\documentclass[useAMS,usenatbib]{mnras}
\usepackage{graphicx}
\usepackage{latexsym}
\usepackage{psfig}
\usepackage{amssymb,amsmath}
\usepackage{subfig}
\usepackage{caption}

\title{The effect of the virial state of molecular clouds on the influence of feedback from massive stars}

\author[James. E. Dale]{James. E. Dale$^{1,2,3}$\thanks{E-mail: dale.james.e@gmail.com(JED)}\\
$^{1}$Centre for Astrophysics Research, University of Hertfordshire, College Lane, Hatfield, AL10 9AB,
UK.\\
$^{2}$Excellence Cluster `Universe', Boltzmannstr. 2, 85748 Garching, Germany.\\
$^{3}$Universit\"{a}ts--Sternwarte M\"{u}nchen, Scheinerstr. 1, 81679 M\"{u}nchen, Germany.}

\begin{document}

\pagerange{\pageref{firstpage}--\pageref{lastpage}} \pubyear{2006}

\maketitle


\def\mnras{MNRAS}
\def\apj{ApJ}
\def\aj{AJ}
\def\aap{A\&A}
\def\apjl{ApJL}
\def\apjs{ApJS}
\def\araa{ARA\&A}
\def\pasp{PASP}
 
\begin{abstract}
A set of Smoothed Particle Hydrodynamics simulations of the influence of photoionising radiation and stellar winds on a series of 10$^{4}$M$_{\odot}$ turbulent molecular clouds with initial virial ratios of 0.7, 1.1, 1.5, 1.9 and 2.3 and initial mean densities of 136, 1135 and 9096\,cm$^{-3}$ are presented. Reductions in star formation efficiency rates are found to be modest, in the range $30\%-50\%$ and to not vary greatly across the parameter space. In no case was star formation entirely terminated over the $\approx3$\,Myr duration of the simulations. The fractions of material unbound by feedback are in the range $20-60\%$, clouds with the lowest escape velocities being the most strongly affected.\\
Leakage of ionised gas leads to the HII regions rapidly becoming underpressured. The destructive effects of ionisation are thus largely not due to thermally--driven expansion of the HII regions, but to momentum transfer by photoevaporation of fresh material. Our simulations have similar global ionisation rates and we show that the effects of feedback upon them can be adequately modelled as a steady injection of momentum, resembling a momentum--conserving wind.\\
\end{abstract}

\section{Introduction}
\indent Star formation is one of the most important subprocesses in the cycling of baryonic matter in the interstellar medium (ISM). Star formation acts as a sink of molecular material, and a source of ionised gas, metal--enriched gas, photons, cosmic rays, momentum and energy. It is one of the primary drivers of the evolution and dynamics of the ISM, and indeed the IGM, via stellar--feedback driven galactic outflows \citep[e.g.][]{2005MNRAS.356..737S,2008A&A...477...79D,2011ApJ...731...41O,2016MNRAS.456.3432G,2016arXiv160605346G}.\\
\indent The various stellar feedback mechanisms influence both the process of star formation itself on the scale of molecular cores, clumps, and clouds, and the larger--scale dynamics of the ISM on the scales of spiral arms or galactic disks. While supernova are almost certainly the most important stellar contribution at these larger scales, many studies have shown that the effects of supernovae are modulated by the environments in which they explode \citep[e.g.][]{2005MNRAS.356..737S,2014A&A...570A..81H,2014MNRAS.443.2092U,2015MNRAS.449.1057G}. The influence of supernovae thus depends on the prior action of other feedback mechanisms, in particular photoionisation \citep[e.g.][]{2008ApJ...682...49W,2015MNRAS.451.2757W}, and/or winds \citep[e.g.][]{2013MNRAS.431.1337R, 2016MNRAS.456..710F} and, especially in very dense systems, radiation pressure \citep[e.g.][]{2010ApJ...710L.142F,2015ApJ...809..187S,2016ApJ...819..137K}.\\
\indent Ionisation also has more subtle large--scale influence owing to the fact that substantial quantities of ionising photons evidently escape from young clusters and their host clouds. This enabled massive stars to contribute to the reionisation of the Universe at high redshifts \citep[e.g.][]{1999ApJ...514..648M,2011MNRAS.412..935P,2015MNRAS.451.1586P}, and supplies photons which help to maintain the thick layers of ionised gas observed to clothe galactic disks, such as the Milky Way's Reynolds Layer \citep[e.g.][]{1995ApJ...448..715R,2004MNRAS.353.1126W,2014MNRAS.440.3027B}.\\
\indent Both supernovae and photoionisation are themselves indirectly influenced by feedback acting on still smaller scales, such as jets and outflows and thermal accretion feedback, since these regulate the rate of formation of the stars which are the sources of ionising photons \citep[e.g.][]{2012ApJ...754...71K,2015MNRAS.450.4035F}. While different types of feedback surely act at different scales, they are thus all connected.\\
\indent In this paper, we examine the effects of ionisation and wind feedback from massive stars acting on molecular cloud scales. The clouds themselves are modelled simply as initially--smooth spheres with Gaussian density profiles and imposed turbulent velocity fields which are allowed to form a few massive stars. At this point each simulation is forked. A control simulation is allowed to continue as before, while in a dual--feedback counterpart, winds and photoionisation from the massive stars are enabled.\\
\indent It is well known that molecular clouds have a range of virial ratios \citep[e.g.][]{2011MNRAS.413.2935D} and our previous work \citep[e.g.][]{2014MNRAS.442..694D} focussed on clouds with initial virial ratios of either 0.7 or 2.3. These values exclude approximately the 20\% most bound and 20\% least bound of the clouds in the \cite{2009ApJ...699.1092H} sample analysed by \cite{2011MNRAS.413.2935D}, and thus roughly the bracket the `typical' 60\% of clouds around the median virial ratio of $\approx1.5$. In this work, we fill in the gap between these two bracketing values of the virial parameter by complementing our original calculations with clouds whose initial value of $\alpha$ is 1.1, 1.5 and 1.9, in order to examine the influence of the clouds' initial energetic states on their response to feedback in a more graduated fashion.\\
\indent The initial density and virial state of a cloud are likely the two most important quantities deciding how fast it will form stars, and we here model clouds with three different initial densities of 136, 1135 or 9096\,cm$^{-3}$. Raising the density decreases the freefall time, which straightforwardly increases the \emph{absolute} rate at which the cloud converts gas to stars, as measured by the star--formation efficiency rate, $SFER$, the rate at which the star formation efficiency increases. However, the star formation efficiency rate \emph{per freefall time}, $SFER_{\rm ff}$ should not depend on the density. Observations \citep[e.g.][]{2007ApJ...654..304K,2011ApJ...727L..12F} suggest that the star formation rate per freefall may be constant over a wide range of densities. Results presented in \cite{2014MNRAS.442..694D} (Figure 16 in that paper) imply that $SFER_{\rm ff}$ indeed correlates weakly with cloud density, but is a function of cloud virial ratio, with the more bound simulations exhibiting $SFER_{\rm ff}$ more than a factor of three higher than the less bound clouds. This reflects the role of turbulence in resisting, preventing or even reversing cloud collapse.\\
\begin{table*}
\begin{tabular}{|l|l|l|l|l|l|l|l|l|l|}
Run&R$_{0}$(pc)&$\alpha_{\rm vir,0}$ & $\langle n(H_{2})_{,0}\rangle$ (cm$^{-3}$) & v$_{\rm RMS,0}$(km s$^{-1}$)& $\langle T_{0}\rangle(K)$&$\langle {\mathcal M}_{0}\rangle$&t$_{\rm ff,0}$ (Myr)&$v_{\rm ESC,0}$(km s$^{-1}$)\\
\hline
r07i&10.0&0.7&136&2.1&52&4.6&2.6&2.9\\
\hline
r07j&5.0&0.7&1135&3.0&30&8.5&0.9&4.1\\
\hline
r11o&5.0&1.1&1135&3.8&30&10.8&0.9&4.1\\
\hline
r15l&5.0&1.5&1135&4.4&30&12.5&0.9&4.1\\
\hline
r15m&2.5&1.5&9096&6.1&18&22.4&0.4&5.8\\
\hline
r19s&5.0&1.9&1135&4.9&30&13.9&0.9&4.1\\
\hline
r19t&2.5&1.9&9096&6.9&18&25.3&0.4&5.8\\
\hline
r23p&2.5&2.3&9096&7.6&18&27.9&0.4&5.8\\
\hline
r23q&5.0&2.3&1135&5.4&30&14.9&0.9&4.1\\
\hline
\end{tabular}
\caption{Initial properties of clouds. The meaning of the columns is as follows: 1: run identifier; 2: initial radius (pc); 3: initial virial parameter; 4: initial mean molecular hydrogen number density (cm$^{-3}$); 5: initial root--mean--square 3D turbulent velocity (km\,s$^{-1}$); 6: initial mean temperature (K); 7: initial mean Mach number; 8: initial freefall time at the mean density (Myr); 9: initial escape velocity (km\,s$^{-1}$).}
\label{tab:sims}
\end{table*}
\begin{figure*}
\includegraphics[width=0.99\textwidth]{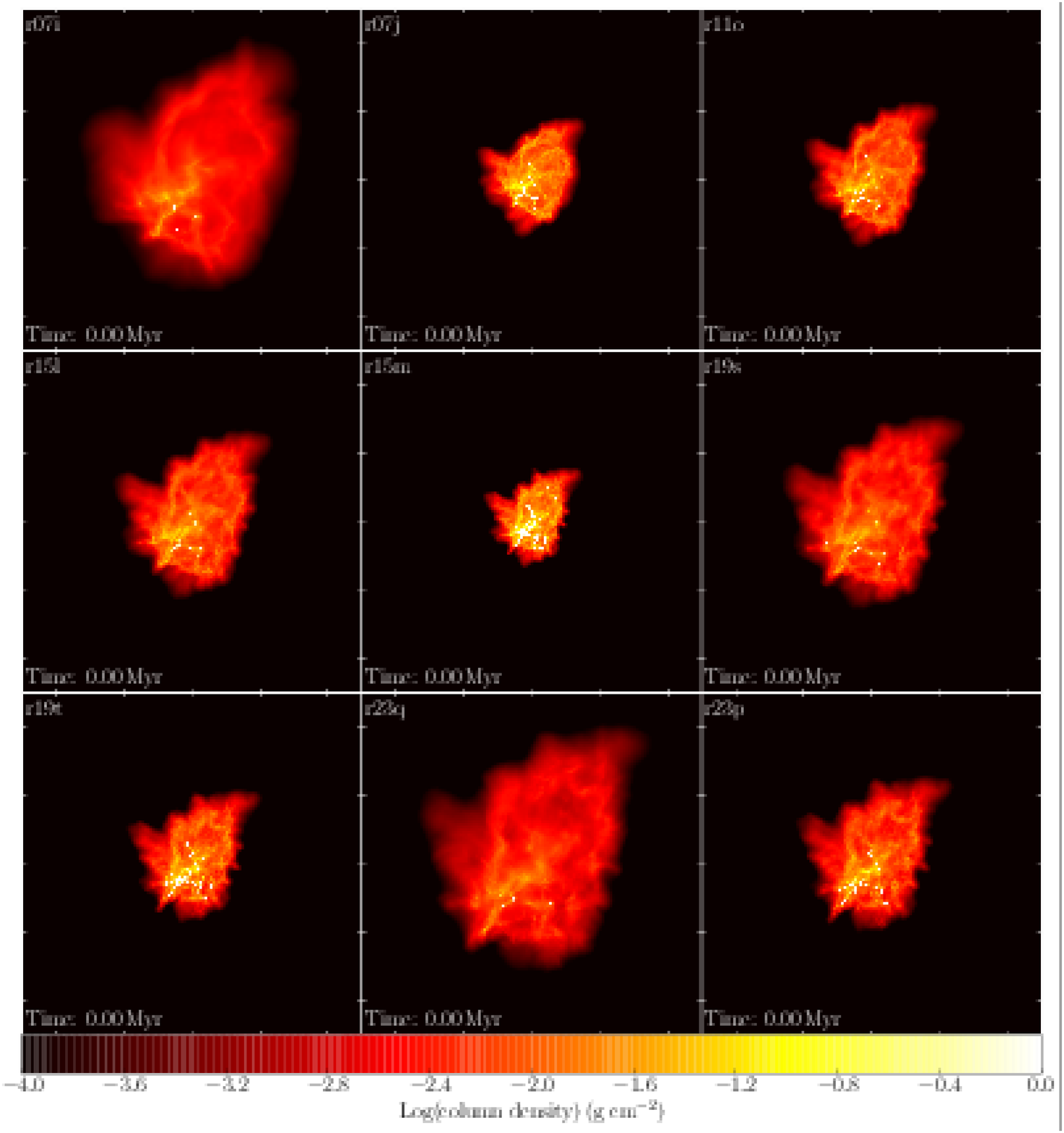}
\caption{Column--density images as viewed along the $z$--axis of the pre--feedback states of the dual--feedback simulations shown at the same 25$\times$25 pc scale. Simulation id's are given in the top left corner and times on the bottom left.}
\label{fig:control_init}
\end{figure*}
\begin{figure*}
\includegraphics[width=0.99\textwidth]{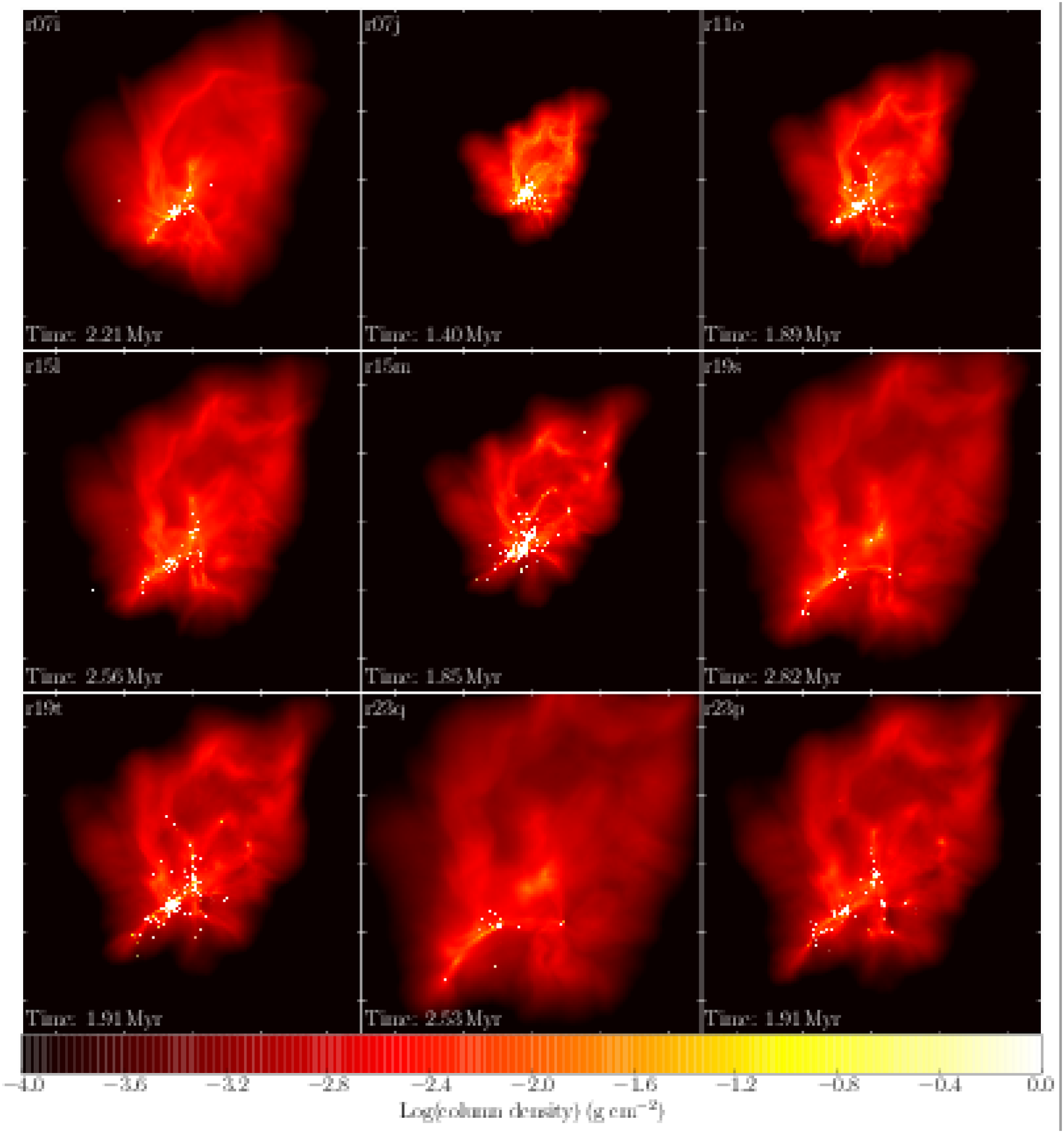}
\caption{Column--density images as viewed along the $z$--axis of the final states of the control simulations shown at the same 25$\times$25 pc scale.  Simulation id's are given in the top left corner and times on the bottom left.}
\label{fig:control_final}
\end{figure*}
\indent We focus only on lower--mass 10$^{4}$M$_{\odot}$ objects to make the simulation set as homogeneous as possible, with the object of modelling a representative sample of clouds of this mass covering a range of virial parameters and densities. The outline for the remainder of this paper is as follows. Section 2 briefly recapitulates our numerical methods, which are identical to previous papers. The results of the simulations are presented in Section 3 and discussed in Section 4. Our conclusions are summarised in Section 5.\\ 
\section{Numerical methods}
We use a hybrid N--body/Smoothed Particle Hydrodynamics (SPH) code based on that described by \cite{1990nmns.work..269B} and implementing point mass sink particles to treat star formation as in \cite{1995MNRAS.277..362B}. We present models of 10$^{4}$M$_{\odot}$ clouds using $10^{6}$ SPH particles, giving a mass resolution of $\approx0.5-1.0$M$_{\odot}$. Sink particle accretion radii are fixed at 0.005 pc and the minimum creation density is $7\times10^{7}$cm$^{-3}$. Sink particle mergers are not permitted and sink--sink gravitational interactions are smoothed at the accretion radius. We model the clouds as initially smooth spheres of gas with a mild Gaussian density profile seeded with a three--dimensional turbulent velocity field. The turbulence is initially divergence--free, satisfying $P(k)\propto k^{-4}$, and decays freely during the simulations.\\
\indent As in our earlier work, the thermodynamics of the cold gas is treated in a simplified fashion using a piecewise barotropic equation of state from \cite{2005MNRAS.359..211L}, defined so that $P = k \rho^{\gamma}$, where
\begin{eqnarray}
\begin{array}{rlrl}
\gamma  &=  0.75  ; & \hfill &\rho \le \rho_1 \\
\gamma  &=  1.0  ; & \rho_1 \le & \rho  \le \rho_2 \\
\gamma  &=  1.4  ; & \hfill \rho_2 \le &\rho \le \rho_3 \\
\gamma  &=  1.0  ; & \hfill &\rho \ge \rho_3, \\
\end{array}
\label{eqn:eos}
\end{eqnarray}
and $\rho_1= 5.5 \times 10^{-19} {\rm g\ cm}^{-3} , \rho_2=5.5 \times10^{-15} {\rm g cm}^{-3} , \rho_3=2 \times 10^{-13} {\rm g\ cm}^{-3}$. At low densities, $\gamma$ is less than unity, implicitly ensuring that the gas in this regime has temperatures in excess of the canonical cloud temperature of $10$K. Dust cooling of the gas is modelled implicitly by the isothermal $\gamma=1.0$ segment, and the $\gamma=1.4$ segment represents the regime where dense collapsing cores become adiabatic. The final isothermal phase of the equation of state is simply in order to allow sink-particle formation to occur. The minimum gas temperature is set to 7.5K, fixing the relation between $\rho$ and $T$. All our simulated clouds have initial average densities $< \rho_{1}$, so that they lie in the line--cooling regime with mean gas temperatures in excess of 10K.\\
\indent The individual initial properties of the simulated clouds are given in Table \ref{tab:sims}. Cloud radii are set to 2.5, 5.0 or 10.0 pc. The magnitudes of the turbulent velocities are normalised to control the virial ratios of the clouds. The gravitational potentials of each one are computed exactly using the SPH code's gravity tree and the velocity fields set to give virial ratios (defined as the simple ratio of total kinetic energy in the centre--of--mass frame to total gravitational binding energy $E_{\rm kin}/|E_{\rm grav}|=\alpha_{\rm vir}=0.7,1.1,1.5,1.9,2.3$). Clouds with $\alpha_{\rm vir}=0.7$ and 2.3 have already been presented in previous papers \citep[][]{2012MNRAS.424..377D,2013MNRAS.430..234D}.\\
\indent Calculations are evolved until three sinks with masses in excess of $20$M$_{\odot}$ have formed at which point they are forked into a control simulation which proceeds as before, and a dual--feedback run with ionisation and wind feedback enabled. Sinks exceeding the mass threshold are assigned ionising photon and wind momentum fluxes as described in \cite{2012MNRAS.424..377D} and \cite{2013MNRAS.436.3430D} respectively. The simulation pairs are then evolved for as close as is practicable to 3 Myr to determine the effects of pre--supernova O--star feedback.\\
\indent The two feedback mechanisms considered -- photoionising radiation and main--sequence stellar winds -- are modelled in the same fashion as in previous papers. A `Str\"omgren volume' technique is used to locate the ionisation fronts by performing reverse--ray--tracing on all gas particles and computing a discretised approximation to the directionally--dependent Str\"omgren integral given in \cite{2007MNRAS.382.1759D} to determine if sufficient ionising photons reach each particle during the current timestep to ionise them (if they are neutral), or to keep them ionised otherwise. Ionised particles are maintained at the canonical 10$^{4}$\,K. If an ionised particle is deprived of photons, it is allowed to recombine on the appropriate timescale, after which it descends a cooling curve. The original algorithm of \cite{2007MNRAS.382.1759D} was updated to allow for the presence of multiple ionising sources, as described in \cite{2012MNRAS.424..377D}. The contribution of the recombination rate of a given intermediate particle to the Str\"omgren integral from a given source to a given target particle is multiplied by the ratio of the ionising flux reaching the intermediate particle from that source to the total flux arriving at that particle from all sources. Several iterations are then required to solve for the complete radiation field.\\
\indent Winds are modelled approximately as described in \cite{2013MNRAS.436.3430D}. Unlike, for example, \cite{2013MNRAS.431.1337R}, we do not inject hot gas into the simulations, since this is technically difficult in SPH. We instead inject only the momentum, or ram--pressure, carried by the stellar winds, by locating the gas particles nearest each massive star which are struck first by large numbers of `momentum packets' emitted by the sources in a Monte Carlo fashion. The quantities of momentum received by each of these particles are then converted to forces which are added to the other forces acting upon them. The \textit{thermal} energies of the gas particles impacted by the winds are not modified. The assumption inherent in this method is that, when the fast--moving stellar winds collide with the ISM, the thermal energy generated by the resulting shocks is radiated away instantaneously. This clearly means that we are modelling the lower limit of the effects of the winds. While this is clearly an oversimplification, more sophisticated simulations \citep[e.g][]{2013MNRAS.431.1337R} have shown that much of the hot gas injected by winds in reality in fact leaks out of the host clouds, advecting the thermal energy along with it. Observational studies of several young star--forming regions by \cite{2014MNRAS.442.2701R} also suggest that only a small fraction of the total injected wind energy is converted into work done on the surrounding clouds, and that much is lost by hydrodynamic leakage.\\
\indent To make it easier to analyse the new simulations and the old synoptically, we adopt a uniform naming convention, where each run is given a four character name. The first (redundant) character is `r', the middle two are digits representing the initial virial ratio of the cloud (07, 11, 15, 19 or 23), and the fourth is an identifying letter preserving the names of the older simulations, and adding new letters for the new runs.\\
\begin{figure*}
\includegraphics[width=0.99\textwidth]{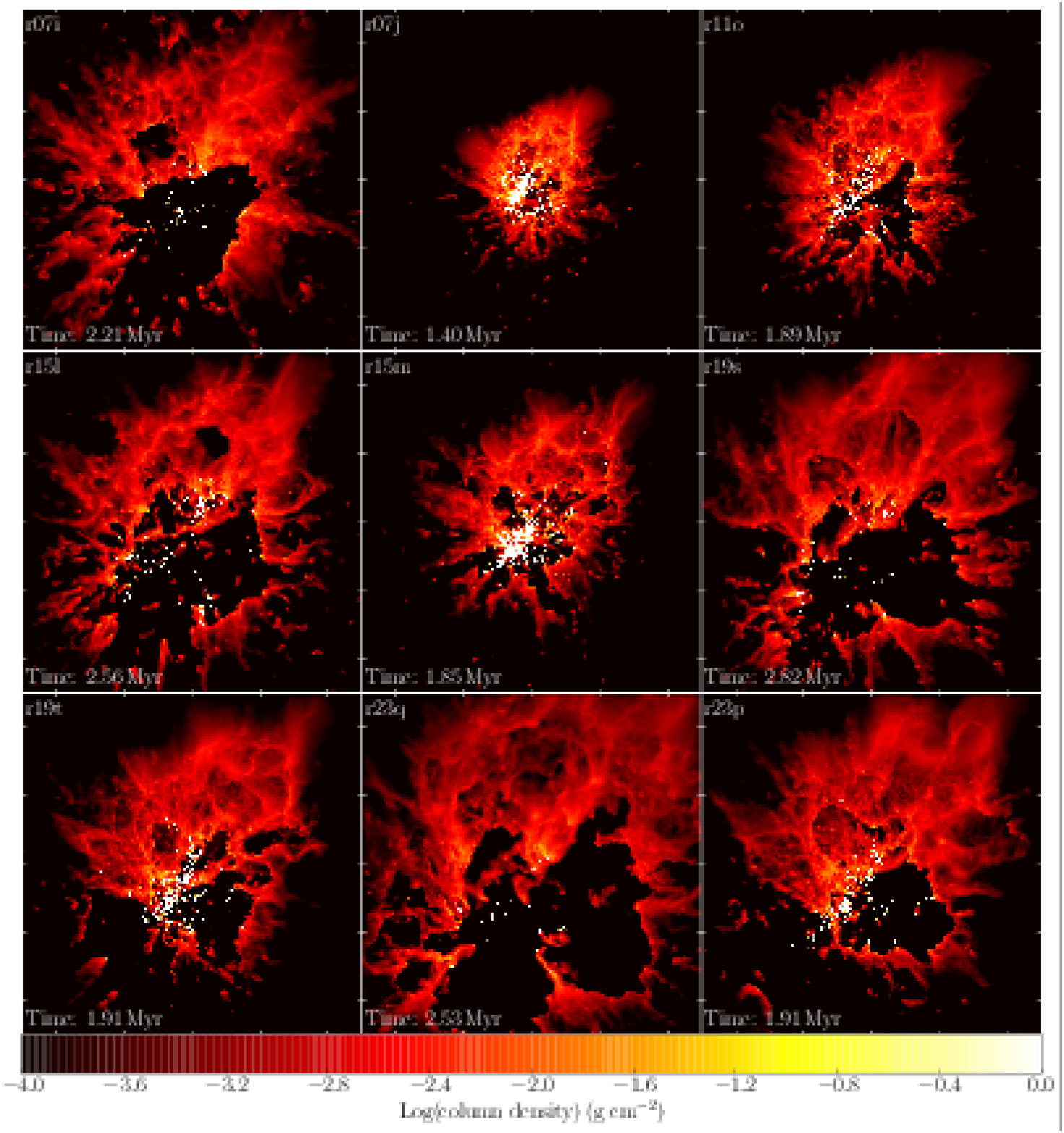}
\caption{Column--density images as viewed along the $z$--axis of the final states of the dual--feedback simulations shown at the same 25$\times$25 pc scale.  Simulation id's are given in the top left corner and times on the bottom left.}
\label{fig:both_final}
\end{figure*}
\begin{figure*}
\captionsetup[subfigure]{labelformat=empty}
\centering
\subfloat[]{\includegraphics[width=0.32\textwidth]{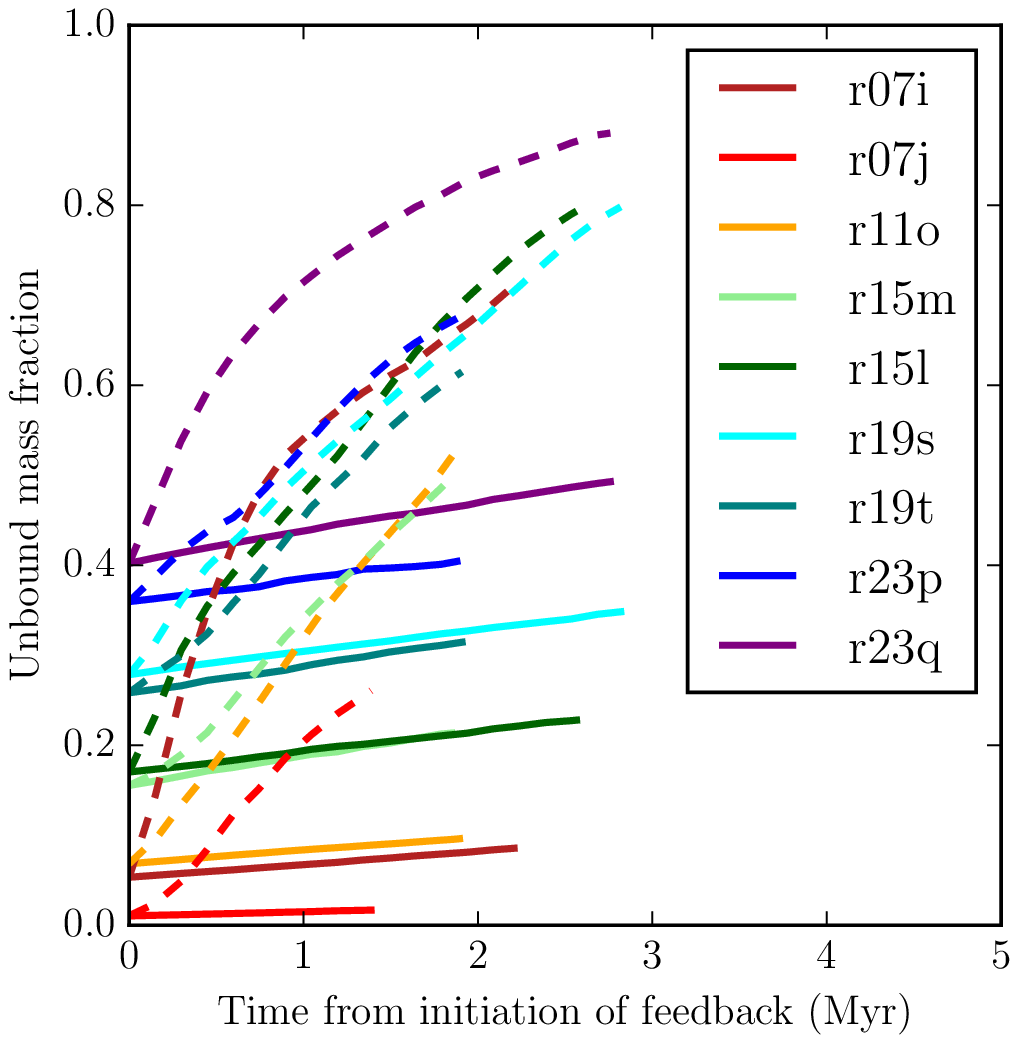}}     
     \hspace{.1in}
\subfloat[]{\includegraphics[width=0.32\textwidth]{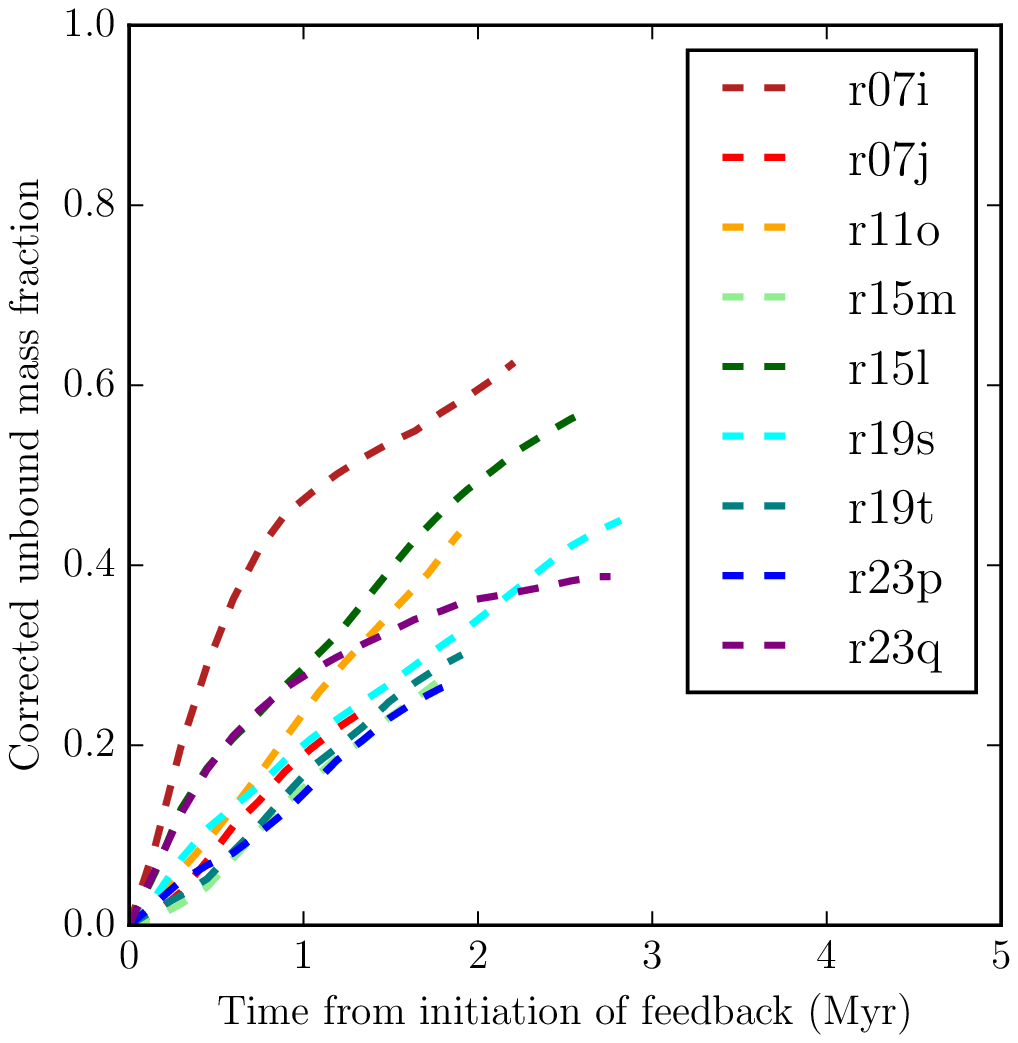}}
     \hspace{.1in}   
\subfloat[]{\includegraphics[width=0.32\textwidth]{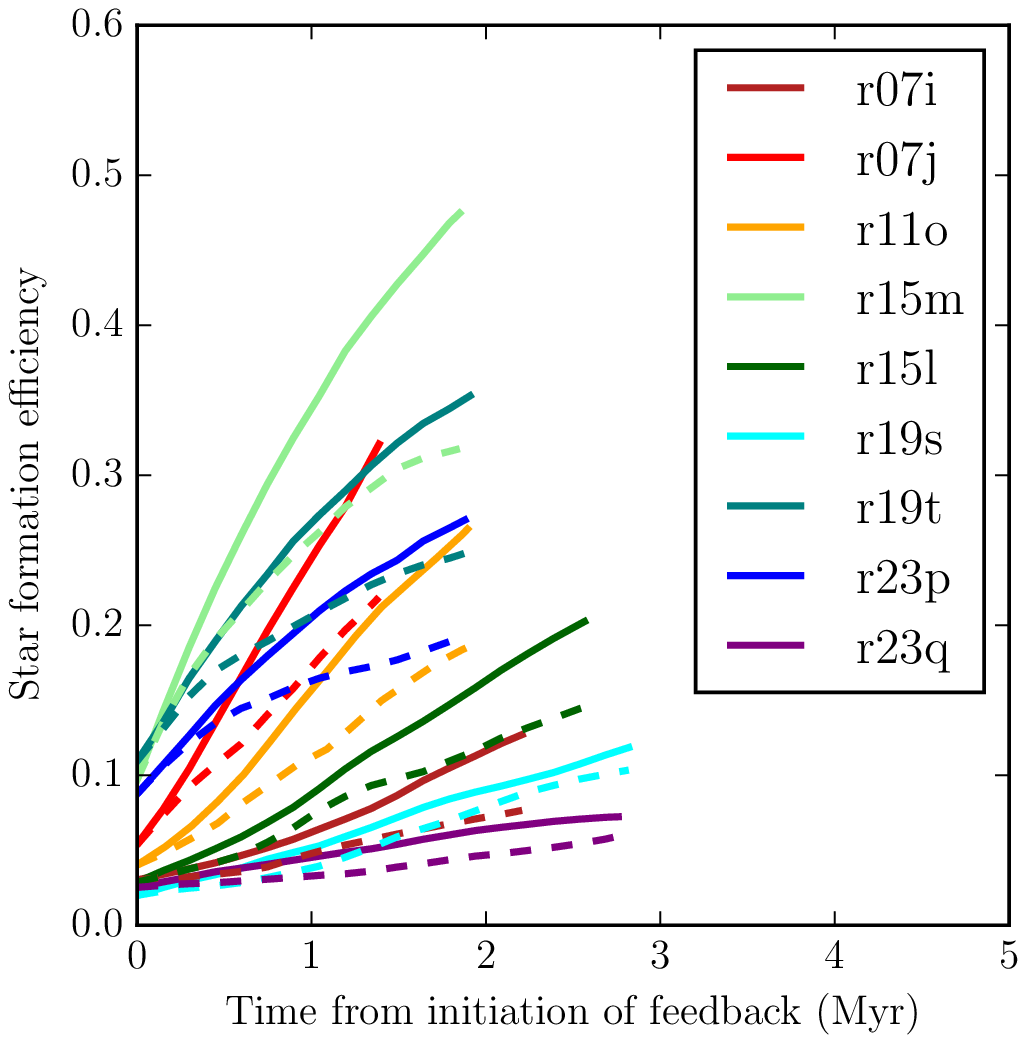}}
     \vspace{.1in}   
\subfloat[]{\includegraphics[width=0.32\textwidth]{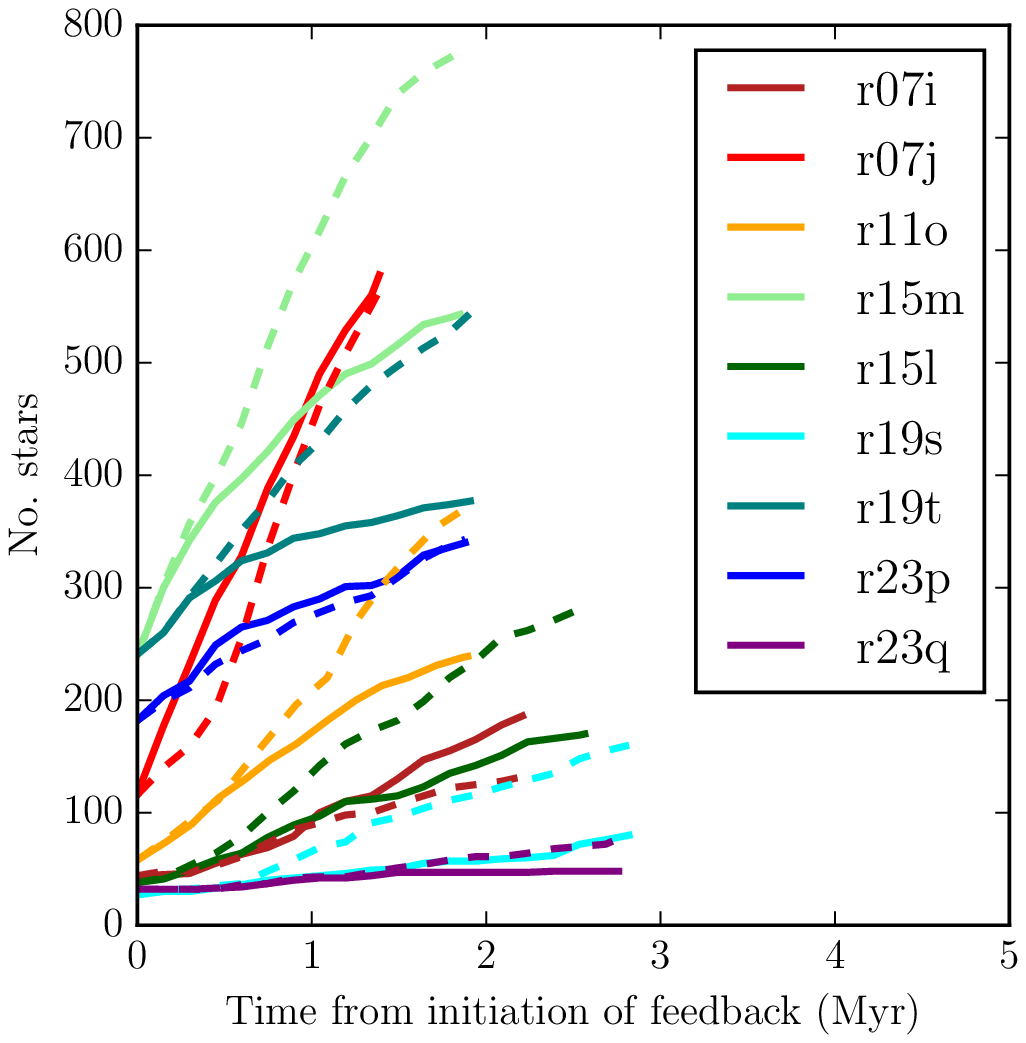}}
     \hspace{.1in}   
\subfloat[]{\includegraphics[width=0.32\textwidth]{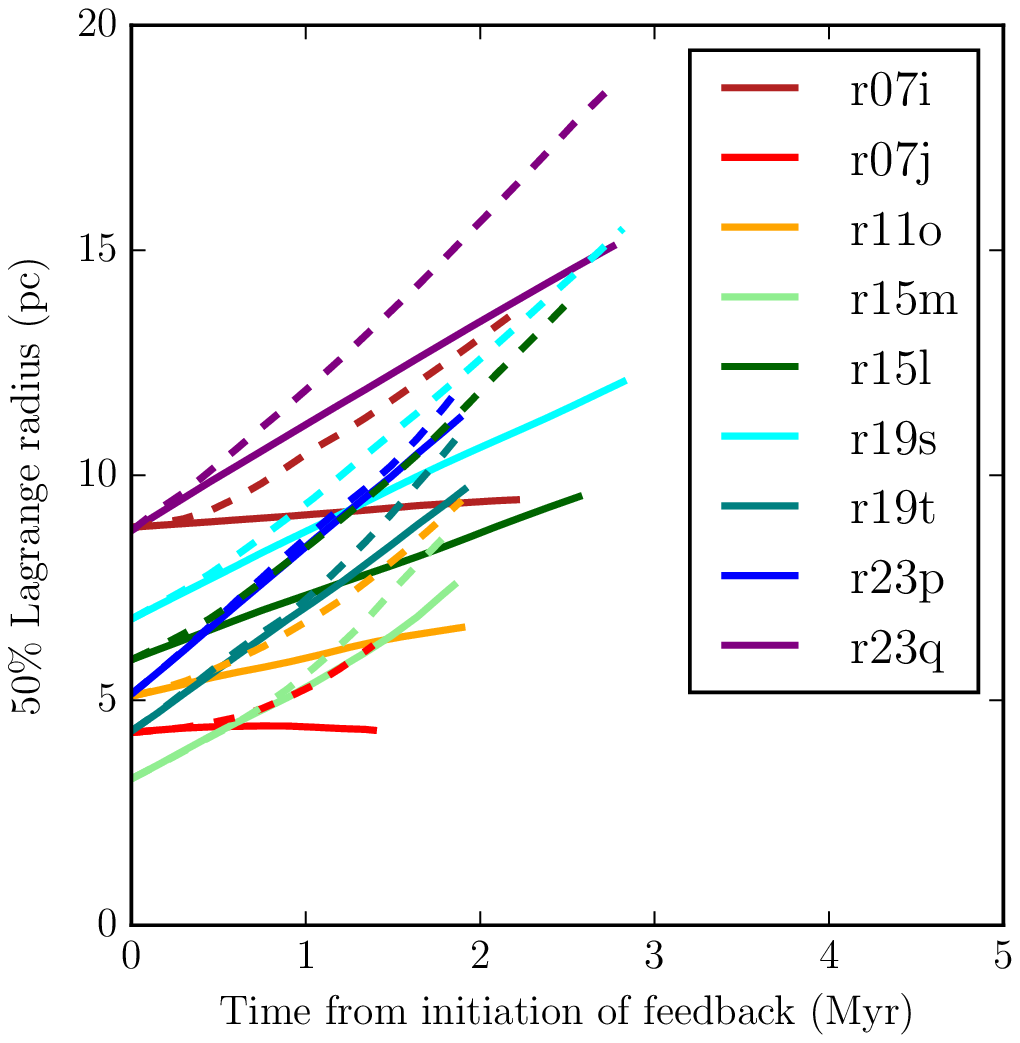}}
    \hspace{.1in}   
\subfloat[]{\includegraphics[width=0.32\textwidth]{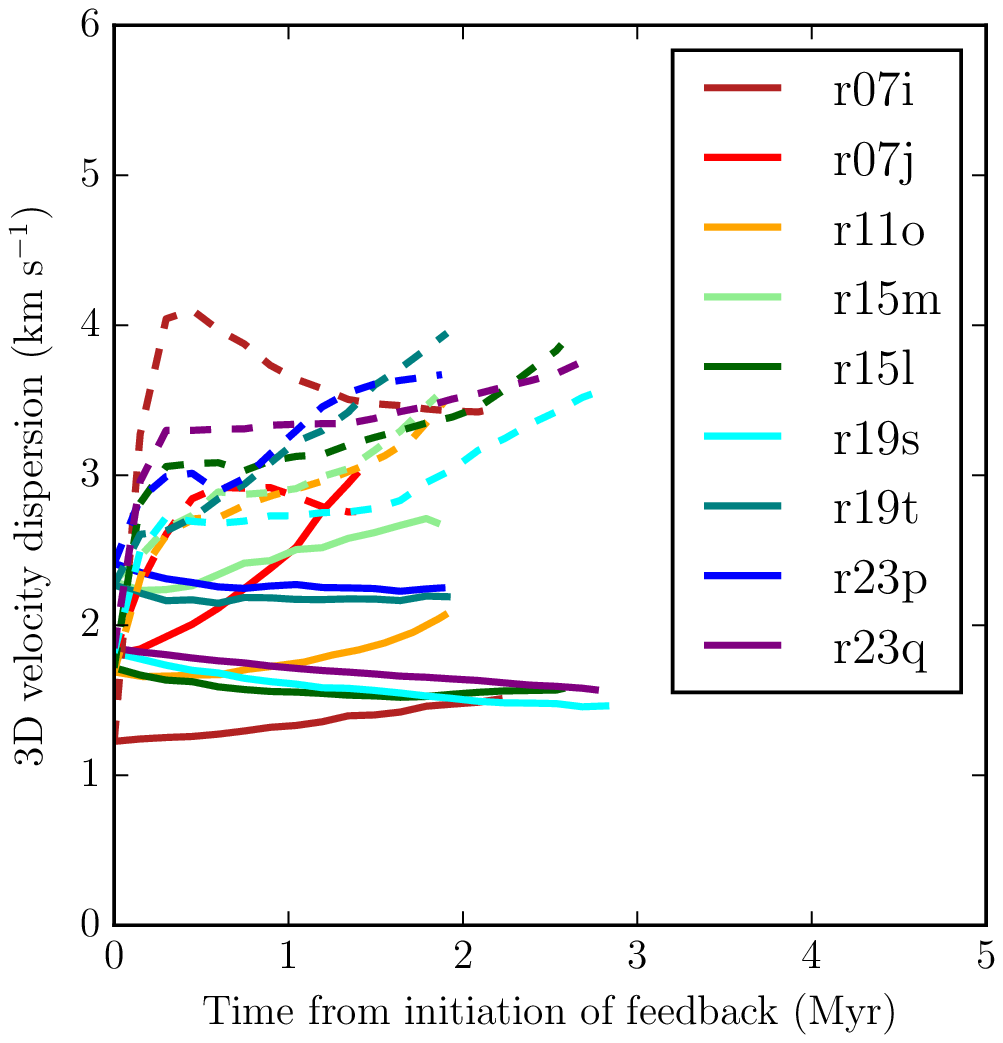}}
\caption{Time evolution of basic diagnostics in all control and dual--feedback simulations: total unbound gas mass (top left), unbound gas mass in feedback runs minus that in control runs (top centre), star formation efficiency (top right), numbers of stars (bottom left), 50$\%$ Lagrange radius (bottom centre), velocity dispersion within 50$\%$ Lagrange radius (bottom right). In all case, solid lines are control simulations, dashed lines are dual--feedback simulations.}
\label{fig:diag}
\end{figure*}
\section{Results}
\subsection{General morphology of clouds}
\indent Figure \ref{fig:control_init} shows column--density images viewed along the $z$--axis of all nine clouds at the same 25$\times$25\,pc scale, with sink particles shown as white dots, just before feedback is initiated, showing the environments that the combined HII regions and wind bubbles first encounter. The most bound clouds are shown at the top left and least bound at the bottom right. The clouds all show to some degree a centrally--condensed filamentary structure, with the main concentrations of stars being around filament junctions or hubs near the centres of clouds and therefore at the deepest parts of their potentials. This configuration resembles that commonly observed in real molecular clouds \citep[e.g.][]{2009ApJ...700.1609M,2012A&A...540L..11S}, and is therefore a good starting point for the study of the influence of feedback.\\
\indent It should be noted, however, that a more realistic starting point would be a cloud formed in a larger (galactic) scale simulation, suitably re--resolved, as has been done recently by \citep[e.g][]{2015MNRAS.446L..46R,2016MNRAS.459.1985S}. While this is undeniable, it is not obvious that arbitrarily rescaling a cloud drawn from a larger simulation in order to study the influence of a particular parameter is an appropriate thing to do. The flexibility offered by the initial conditions used here therefore compensates for their simplicity.\\
\indent Figure \ref{fig:control_final} shows the end results of the control--fork simulations, where feedback is neglected. In general, most of the clouds have expended to some degree, owing to the fact that most have them have virial ratios in excess of unity. It is also apparent that the clouds' central regions have collapsed, resulting in more sharply--defined filaments, and greater concentrations of both stars and gas towards the cloud centres. This is largely a result of the more rapid draining of turbulent energy from the denser central regions of the clouds, as discussed in \cite{2013MNRAS.430..234D}.\\
\indent Figure \ref{fig:both_final} shows the end states of the dual--feedback simulations. Note that Figures \ref{fig:control_init}, \ref{fig:control_final} and \ref{fig:both_final} show all the clouds at the same 25$\times$25\,pc scale, with the same column--density colour scheme. Comparison of Figures \ref{fig:control_final} and \ref{fig:both_final} shows that feedback has clearly had a profound effect, both on the structure of the gas and on the distribution of the stars, in all simulations. Many of the clouds feature well--cleared but irregularly--shaped bubbles. The degree to which the clouds are cleared out appears to vary with the cloud size, in that the larger clouds appear to have had larger fractions of their volumes evacuated by a few coherent bubbles, whereas the smaller clouds, particularly Runs r07j and r11o and r15m, exhibit a jumble of smaller bubbles separated by surviving lanes of cold gas. Since these clouds are all the same mass, smaller size implies higher density, and it therefore appears that the denser clouds are more difficult for feedback to clear effectively.\\
\indent Comparing Figures \ref{fig:control_final} and \ref{fig:both_final} also gives the general impression that the feedback calculations generally have stars spread out over a larger fraction of the cloud volumes, and less condensed in clusters. This is broadly consistent with the conclusion drawn in \cite{2013MNRAS.431.1062D} that feedback \textit{redistributes} star formation. We will return to this issue in detail in a companion paper.\\
\subsection{Basic diagnostic properties}
\indent In Figure \ref{fig:diag} we plot the time--evolution of a series of basic global diagnostic quantities for all the models, using a rainbow colour scheme where the more strongly bound clouds are towards the red end of the spectrum and the less bound towards the violet end, as an aid to interpretation. Solid lines represent control simulations (where appropriate) and dashed lines dual--feedback simulations.\\
\indent The first two panels are concerned with the mass unbound from the clouds. The first panel shows for both control and dual--feedback simulations the total quantity of matter with positive energy in the cloud centre--of--mass frame as a function of time. Note that in all control simulations, some non--zero quantity of mass is always unbound. The amount of such material increases with virial ratio, as expected, but in a given simulation, does not increase very much with time over the timescales considered here, even in the most strongly unbound and rapidly expanding clouds.\\
\indent Turning to the dual feedback simulations, clearly substantial amounts of additional mass are unbound by feedback, but no clouds are destroyed outright. In the second panel of Figure \ref{fig:diag}, we plot the unbound mass fractions in the dual--feedback calculations `corrected' by subtracting at each timestep the fraction of material unbound in the corresponding control simulation. This gives an idea of how much extra material is expelled from the clouds by feedback, on top of what is already unbound due to the remnant turbulent velocity field and the cloud expansion driven by it. We see that 30--60$\%$ of cloud--mass is expelled by this definition, increasing to 40--80$\%$ if the lines were extrapolated to 3Myr. Since the lines of different colours are well mixed, there is no apparent correlation between the virial states of the clouds and how much harm is done to them by feedback.\\
\indent The third panel of Figure \ref{fig:diag} depicts the increase in star formation efficiency with time for all runs. The star formation efficiencies of the simulations span values between 10 and 50$\%$ and some of the control simulations would likely achieve efficiencies of $\approx80\%$ if they were allowed to continue to 3Myr. In all simulations, feedback reduces the star formation rates and efficiencies, but not by large factors, in no case by more than a factor of two. There is once again no apparent correlation with the virial parameter.\\
\indent The fourth panel of Figure \ref{fig:diag} shows the increase in the numbers of stars with time for all calculations. Not only is there no connection between the number of stars each cloud produces and its virial state, but the effect of feedback on the number of stars formed can be in either direction, reflecting the multiple roles of feedback discussed in previous papers \citep[e.g][]{2013MNRAS.431.1062D} of triggering, aborting and redistributing star formation. Recall that the effect of feedback on the total stellar \emph{mass} is \emph{always} negative.\\
\begin{figure*}
\captionsetup[subfigure]{labelformat=empty}
\centering
\subfloat[]{\includegraphics[width=0.33\textwidth]{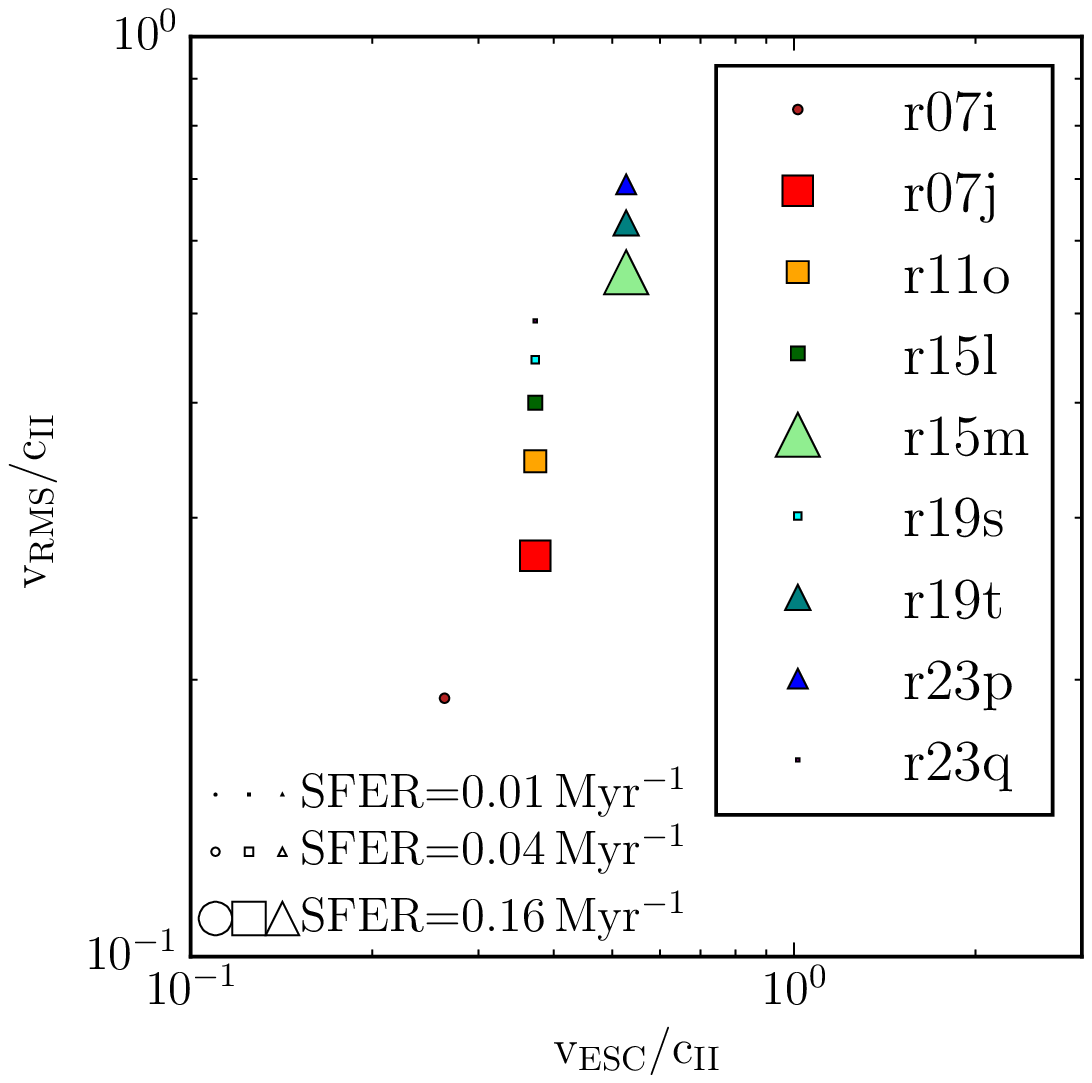}}     
     \hspace{.0in}
\subfloat[]{\includegraphics[width=0.33\textwidth]{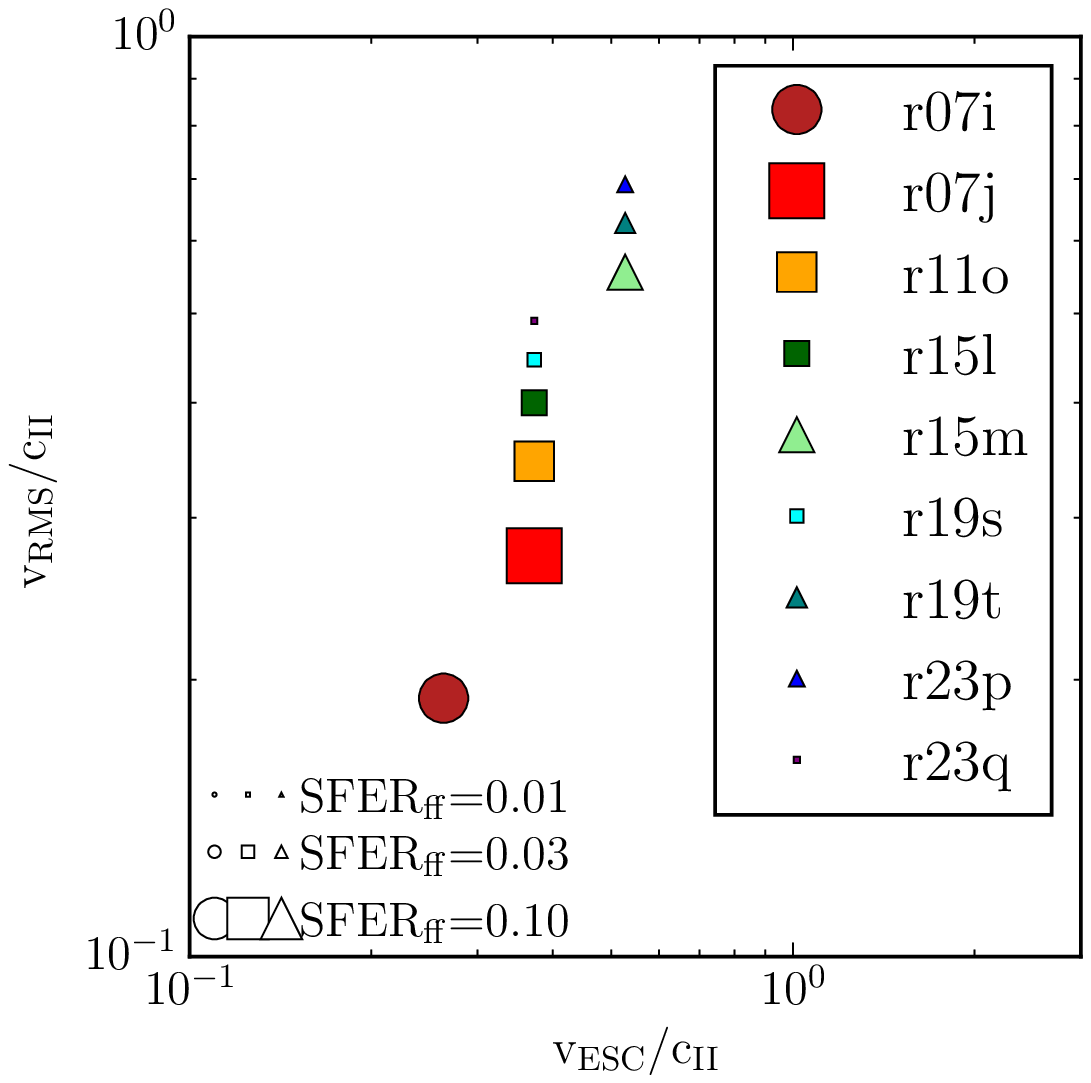}}
     \hspace{.0in}   
\subfloat[]{\includegraphics[width=0.33\textwidth]{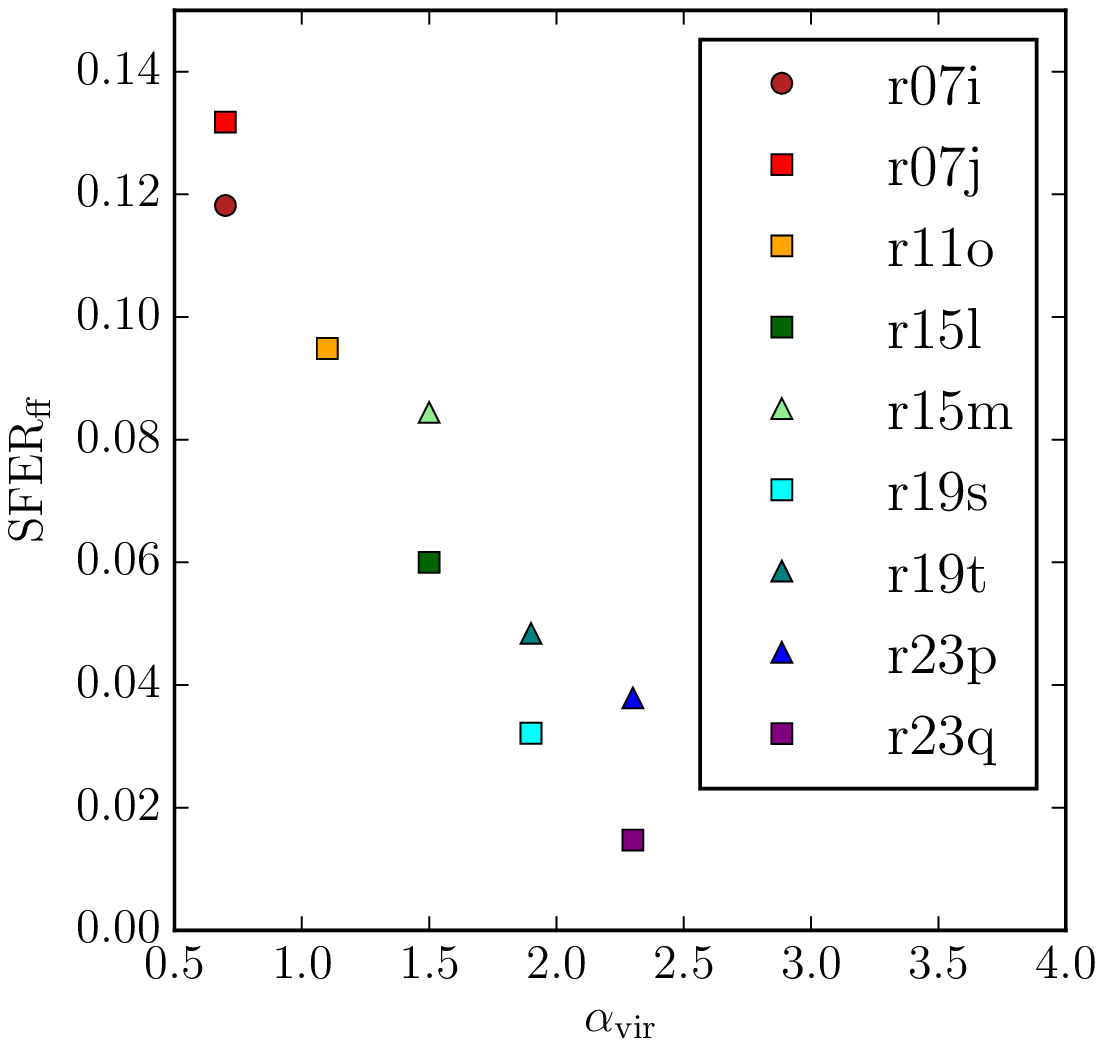}}
\caption{Star formation efficiency rates in the control simulations per Myr (left panel) and per freefall time (centre panel) plotted by symbol size across the [escape velocity:turbulent velocity dispersion] parameter space, and against cloud initial virial ratio (right panel). Rates are computed for the control simulations between the point where the simulations are forked (the initiation of feedback in the dual--feedback simulations) and the ends of the simulations..}
\label{fig:ctrl_param}
\end{figure*}
\begin{figure*}
\captionsetup[subfigure]{labelformat=empty}
\centering
\subfloat[]{\includegraphics[width=0.33\textwidth]{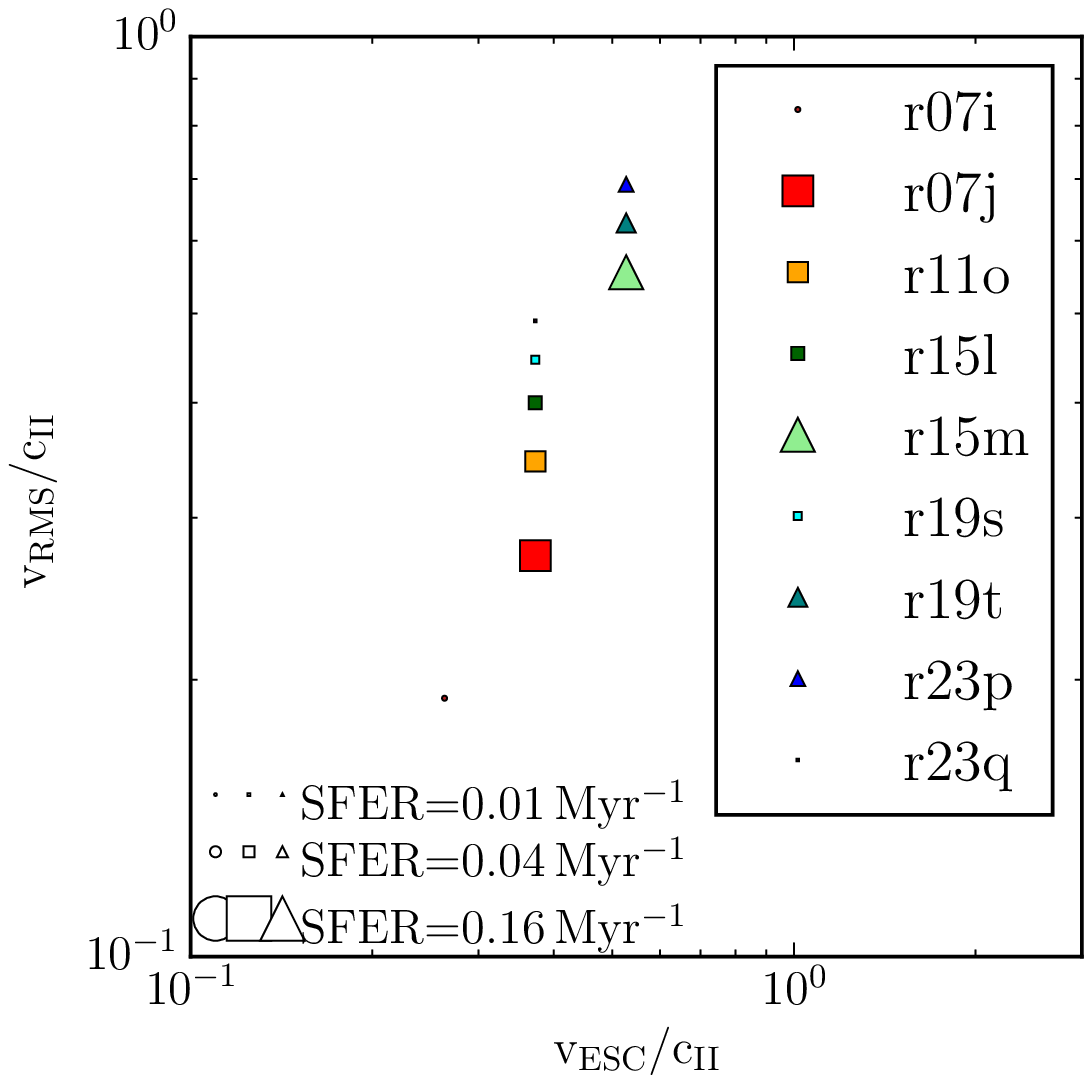}}     
     \hspace{.0in}
\subfloat[]{\includegraphics[width=0.33\textwidth]{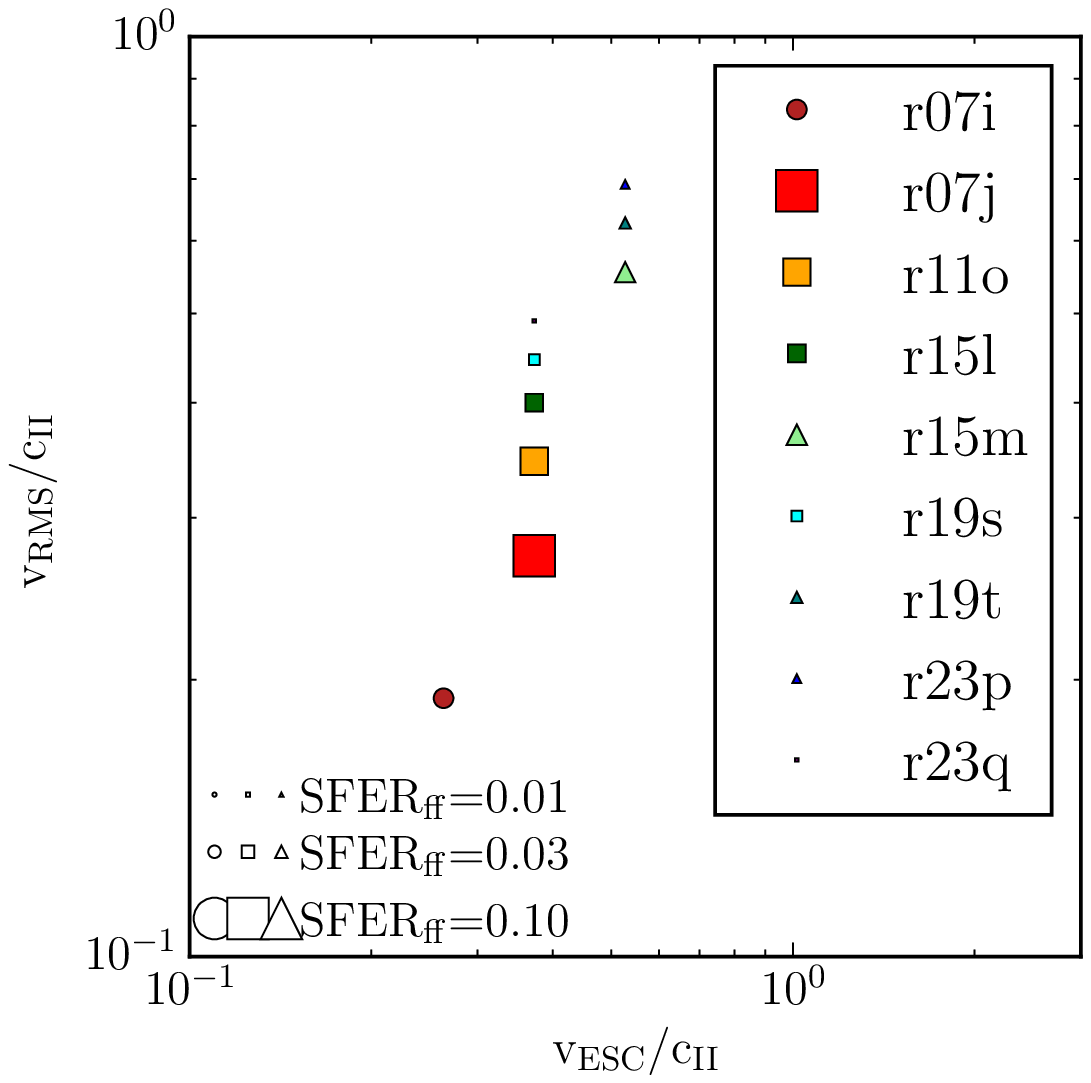}}
     \hspace{.0in}   
\subfloat[]{\includegraphics[width=0.33\textwidth]{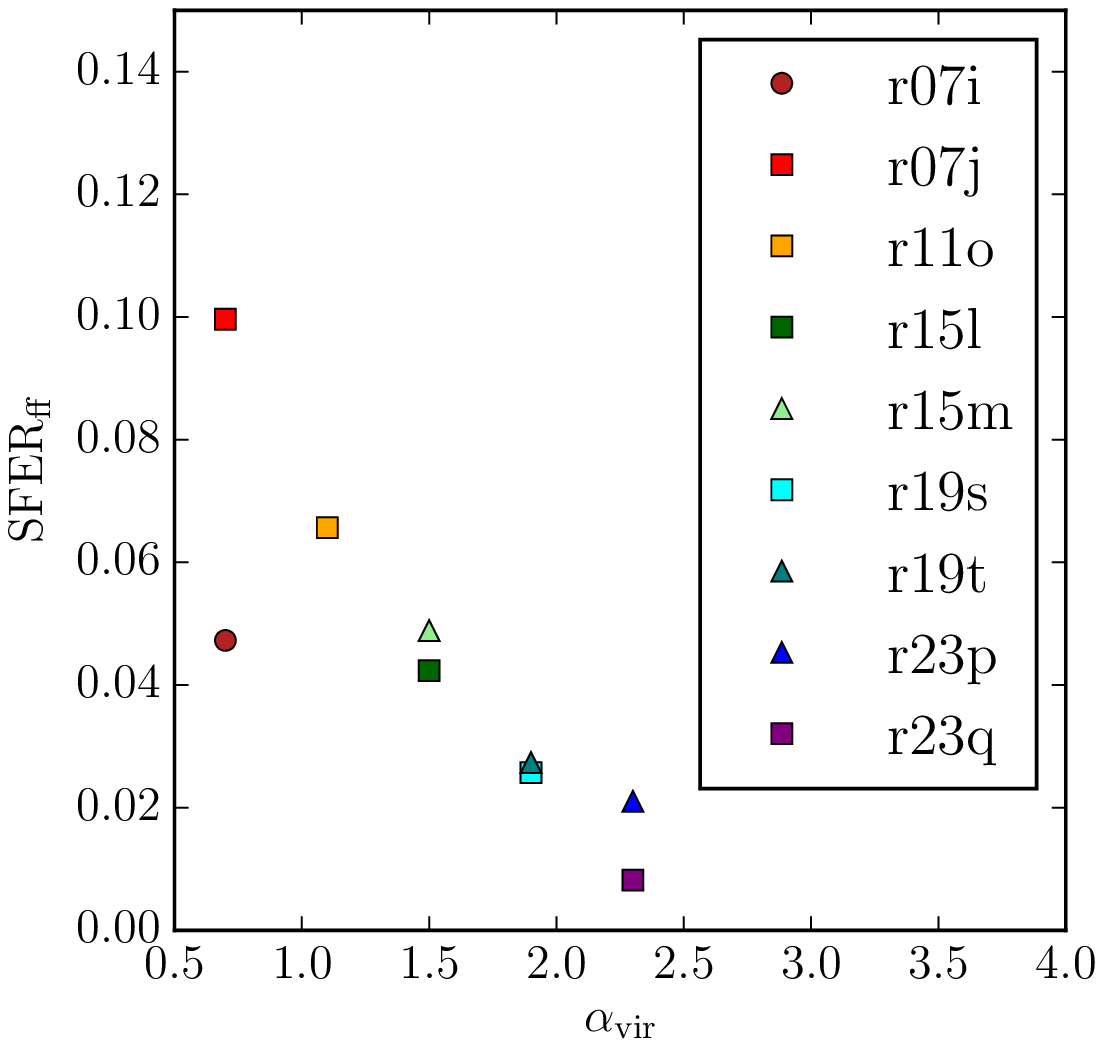}}
\caption{Star formation efficiency rates in the dual--feedback runs per Myr (left panel) and per freefall time (centre panel) plotted by symbol size across the [escape velocity:turbulent velocity dispersion] parameter space, and against cloud initial virial ratio (right panel). Rates are computed for the dual--feedback  simulations between the initiation of feedback and the ends of the simulations.}
\label{fig:df_param}
\end{figure*}
\begin{figure*}
\captionsetup[subfigure]{labelformat=empty}
\centering
\subfloat[]{\includegraphics[width=0.33\textwidth]{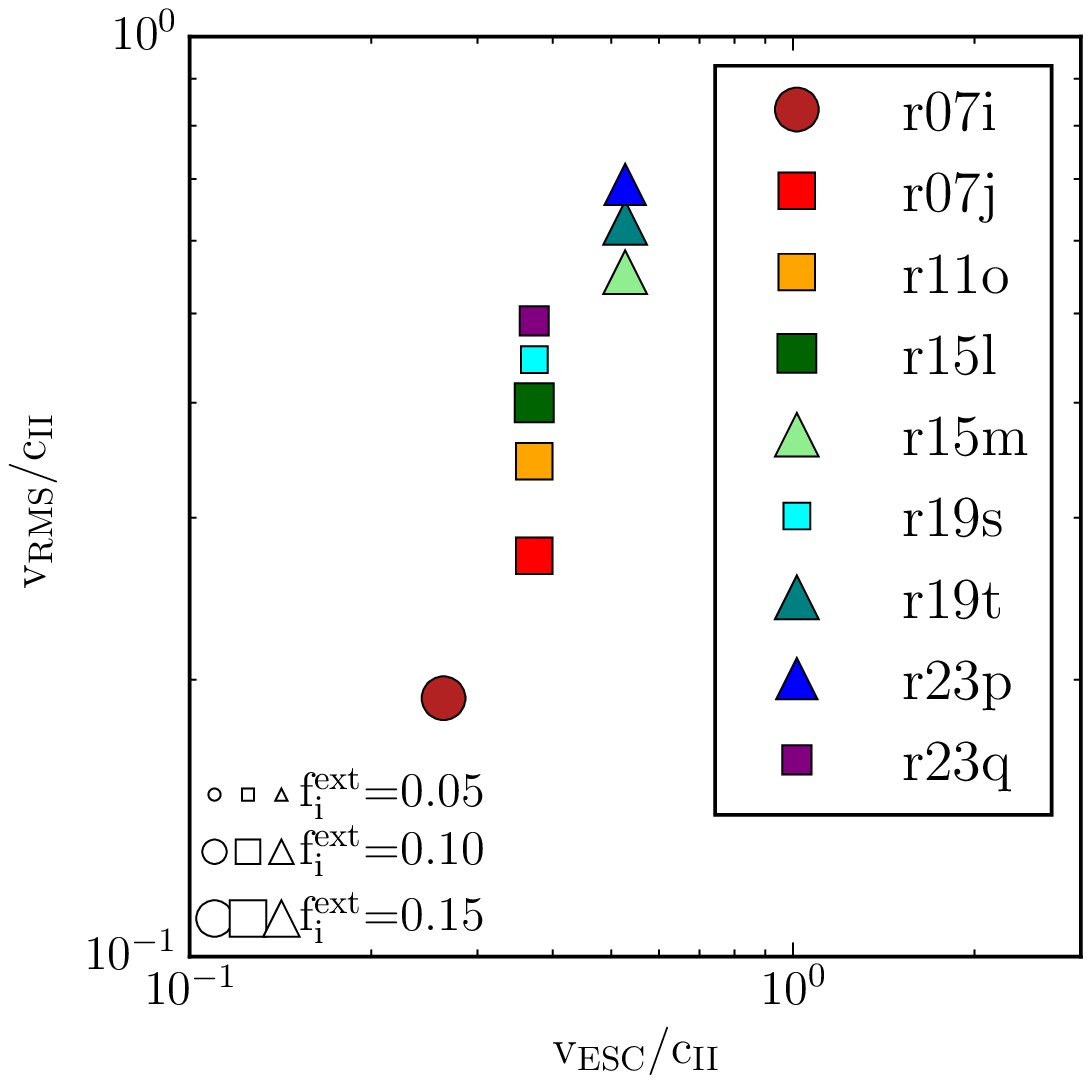}}     
     \hspace{.0in}
\subfloat[]{\includegraphics[width=0.33\textwidth]{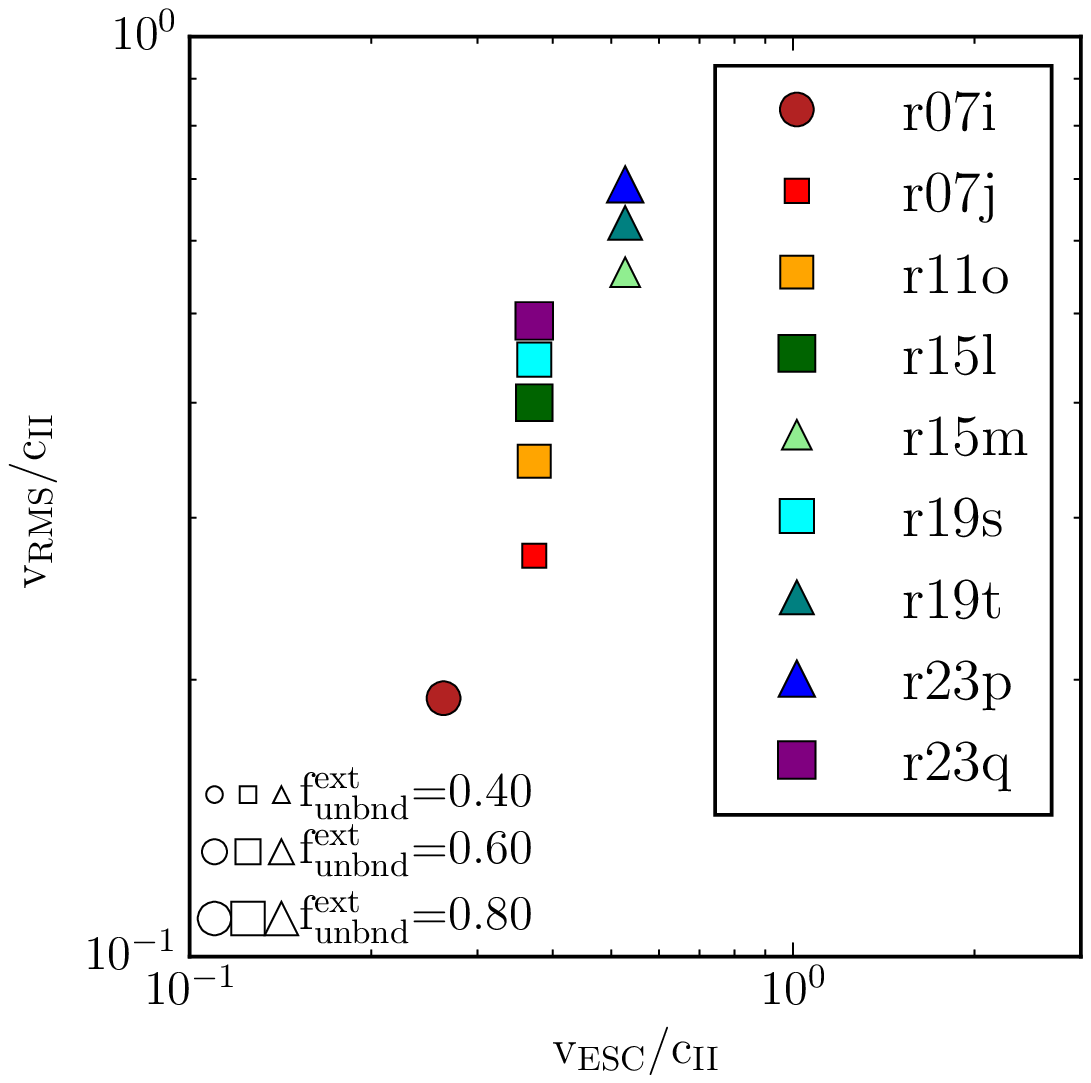}}
     \hspace{.0in}   
\subfloat[]{\includegraphics[width=0.33\textwidth]{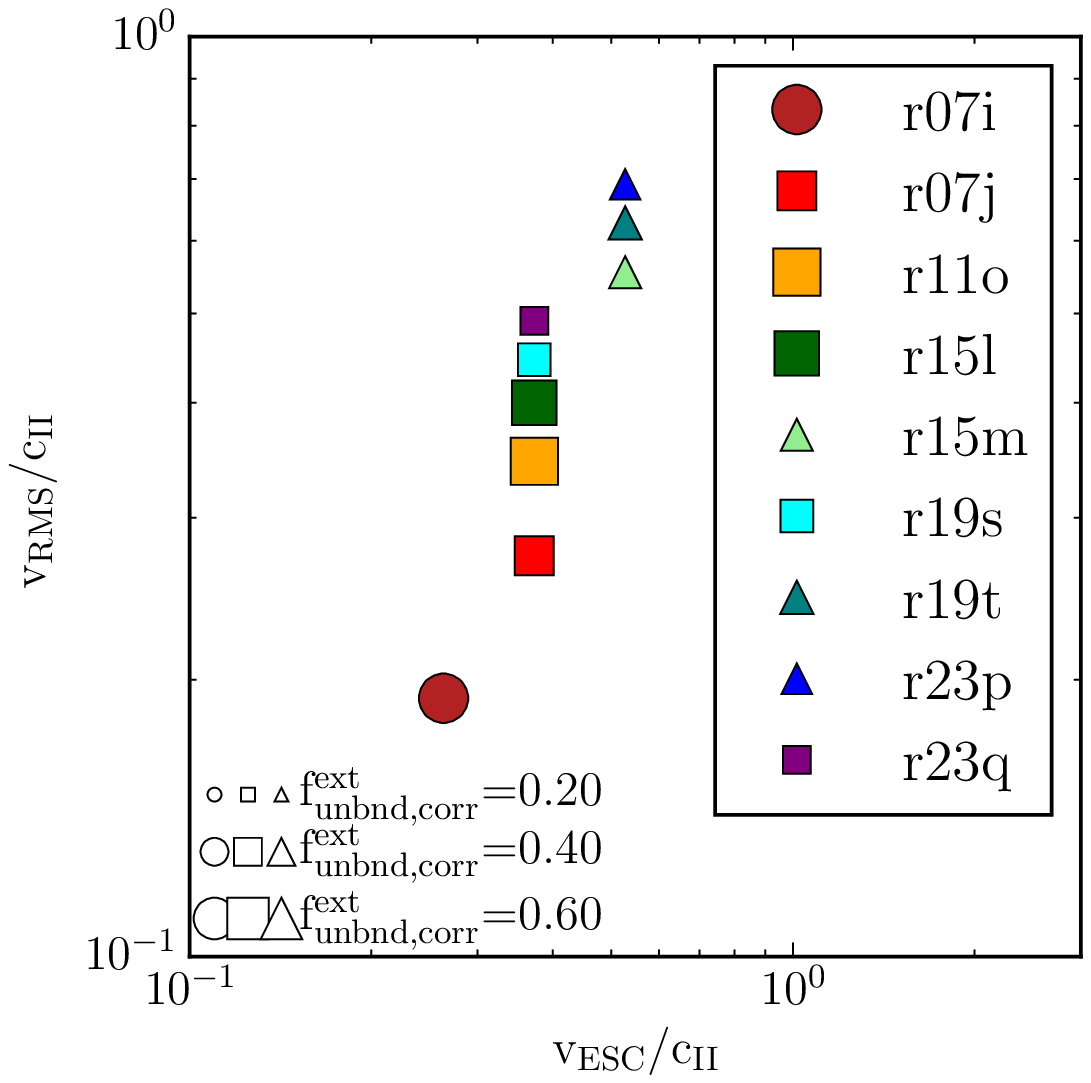}}
\caption{Ionisation fractions (left), total unbound mass (centre) and unbound mass in the dual--feedback simulation minus that in the control simulation (right) plotted in [escape velocity:turbulent velocity] parameter space. All quantities have been extrapolated to a common time of 3\,Myr after the initiation of feedback.}
\label{fig:unbnd_param}
\end{figure*}
\indent The fifth and sixth panels of Figure \ref{fig:diag} show the evolution of the 50$\%$ Lagrange radii and the three--dimensional velocity dispersion within this volume. All the clouds save r07j expand to some degree, with expansion in the control simulations being driven purely by the fading turbulent velocity fields, so that the more unbound clouds expand by larger factors. The expansion is, not surprisingly, substantially faster in the dual--feedback simulations. The velocity dispersions of the more bound control simulations grow with time for the more bound clouds, reflecting the gravitational collapse in the cloud cores, which also causes the star formation rates in these clouds to hold steady or accelerate somewhat. In the less bound clouds, the velocity dispersion declines as the clouds' turbulence dies away and decreasing density leads to a steady diminution in star formation activity. In the dual--feedback simulations, the velocity dispersion increases very sharply as feedback is enabled, and then plateaus or even declines as the HII region expansion is slowed by mass--loading.\\
\subsection{Variations across parameter space}
\indent It is instructive to plot the values at common epochs of some of the quantities discussed above in the [escape velocity:turbulent velocity dispersion] parameter space in which these simulations were conceived, and against the initial virial ratio of the clouds, which we do in Figures \ref{fig:ctrl_param}, \ref{fig:df_param} and \ref{fig:unbnd_param}. The escape velocity on the $x$--axis and the initial turbulent velocity dispersion on the $y$--axis are both normalised by the ionised sound speed of 11.3\,km\,s$^{-1}$. The different plot symbols represent different cloud initial densities as a reminder of the variation of this parameter -- circles denote $\langle n(H_{2})_{,0}\rangle$=136\,cm$^{-3}$, squares $\langle n(H_{2})_{,0}\rangle$=1135\,cm$^{-3}$ and squares $\langle n(H_{2})_{,0}\rangle$=9096\,cm$^{-3}$.\\
\indent In Figure \ref{fig:ctrl_param}, we plot by symbol size and for the control simulations, the mean star formation rate per Myr (left panel) and per freefall time (centre panel) from the point where the simulations are forked until the ends of the control runs over the parameter space. The right panel instead plots $SFER_{\rm ff}$ directly against the cloud initial virial ratios.\\
\indent The star formation efficiency rates per Myr and per freefall time vary by roughly an order of magnitude amongst the simulations. In general, the clouds with the highest densities and the lowest virial ratios form stars fastest. The relation between $SFER_{\rm ff}$ and $\alpha$ appears very close to a linear decline, albeit with some scatter.\\
\indent Figure \ref{fig:df_param} depicts the same quantities shown using the same scales, but for the dual--feedback simulations in the period between the initiation of feedback and the ends of the simulations. In general, the same trends in these quantities across the parameter space and with the virial ratio are repeated here, except that the star formation efficiency rates, measured either in absolute times or per freefall times, are scaled down by factors of $\approx30-50\%$. In the absence of feedback, most simulations achieve $SFER_{\rm ff}$ of $>5\%$ and two simulations push this over 10\%. None of the dual--feedback simulations reach such high $SFER_{\rm ff}$ and all but two (Runs r07j and r11o) exhibit $SFER_{\rm ff}<5\%$. This demonstrates that, in these smaller clouds, it is possible for the combined effects of winds and ionisation alone to reduce star formation efficiency rates per freefall time to values factors of only a few larger than those observed. Similar values of $SFER_{\rm ff}$ were recently reported by \cite{2015MNRAS.450.4035F}, who modelled the combined effects of driven turbulence, jets and magnetic fields.\\
\indent As shown by Figure \ref{fig:diag}, in no case is star formation \textit{in whole clouds} terminated, so that these clouds would continue towards ever larger star formation efficiencies absent some other effect which turns star formation off. However, feedback redistributes star formation to the peripheries of the systems and many of the simulations exhibit stellar clusters/subclusters which are devoid of gas and which are not accreting from the remains of their host clouds, so star formation within them has been \textit{locally terminated}. This is an encouraging sign from the point of view of reproducing observations, since observations of very young massive clusters empty of gas and without ongoing star formation abound \citep[e.g.][]{2014MNRAS.445..378B, 2015MNRAS.449.1106H}, and such systems appear to become gas--free on timescales of a few Myr. It should be noted, however, that most of the clusters studied in the cited works are much more massive than those produced here, which are at most a few$\times10^{3}$M$_{\odot}$.\\
\indent In Figure \ref{fig:unbnd_param}, we plot in the same parameter space the ionisation fractions, unbound mass fractions, and the corrected unbound mass fractions. All quantities are extrapolated to a common time of 3\,Myr after the initiation of feedback. We see that there is very little variation in the first of these, with $\approx10\%$ of the mass of all clouds being ionised. The variation in the total unbound mass is also modest, with all simulations finishing with unbound mass fractions of 40--80\%, and hints that, for a given escape velocity, clouds with higher turbulent velocities finish with larger unbound mass fractions. The corrected unbound mass fractions show a similar degree of variation and there is a trend that clouds with higher escape velocities suffer less damage, as we found in \cite{2014MNRAS.442..694D}.\\
\section{Discussion}
\begin{figure*}
\captionsetup[subfigure]{labelformat=empty}
\centering
\subfloat[Run r07i, total (left), cold gas (centre), and hot gas momentum (right). Solid lines: control run. Dashed lines: Dual--feedback run.]{\includegraphics[width=0.98\textwidth]{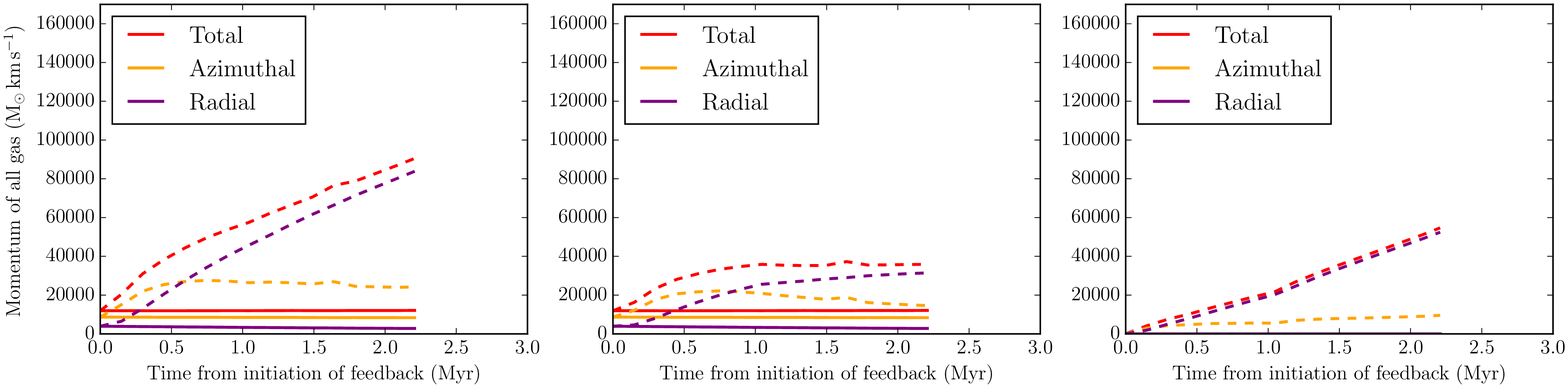}}     
     \vspace{.1in}
\subfloat[Run r15l, total (left), cold gas (centre), and hot gas momentum (right). Solid lines: control run. Dashed lines: Dual--feedback run.]{\includegraphics[width=0.98\textwidth]{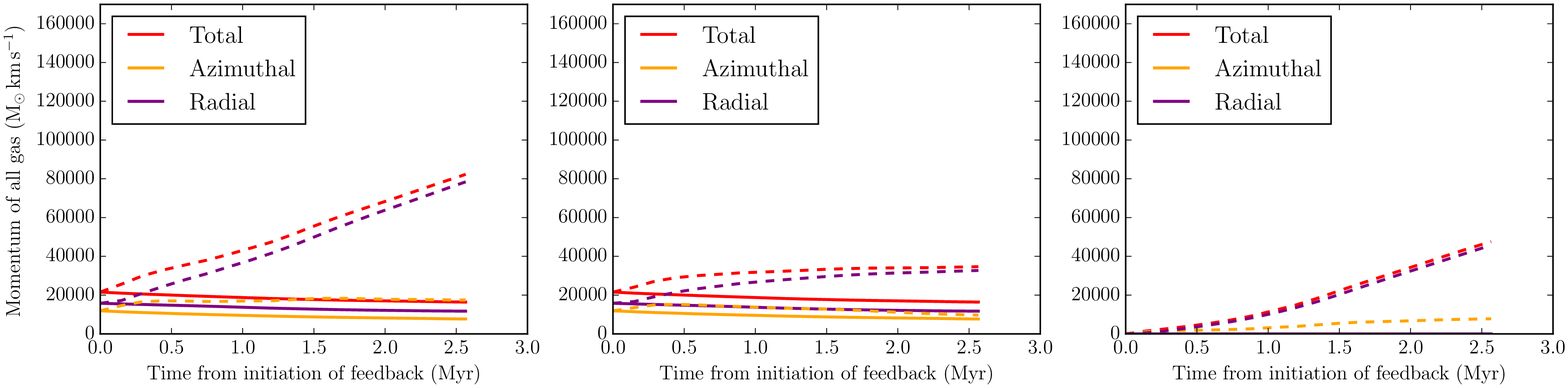}}
     \vspace{.1in}   
\subfloat[Run r23q, total (left), cold gas (centre), and hot gas momentum (right). Solid lines: control run. Dashed lines: Dual--feedback run.]{\includegraphics[width=0.98\textwidth]{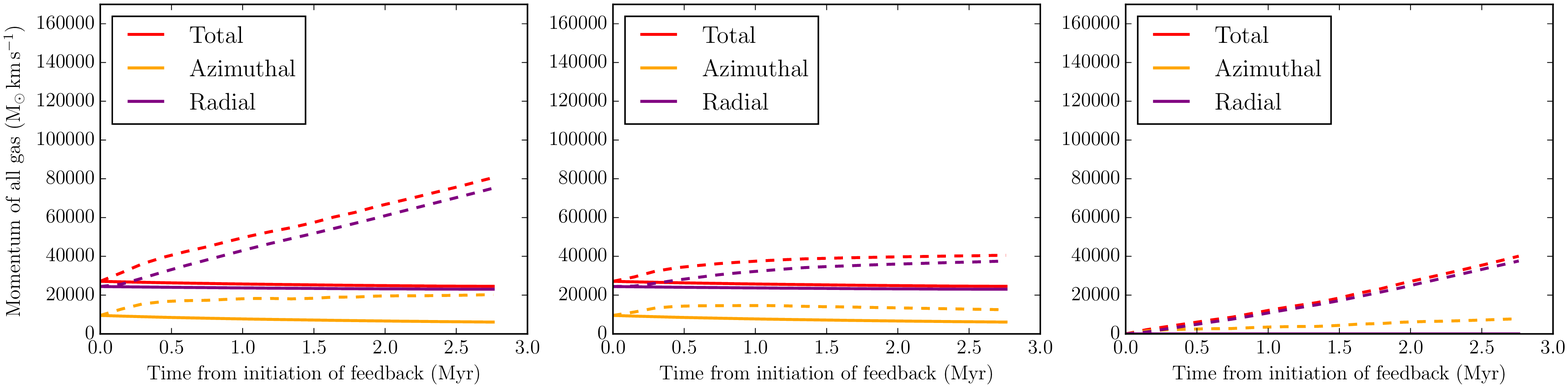}}
\caption{Time evolution of total (red lines), radial (purple lines) and azimuthal (orange lines) momenta in three representative simulations.}
\label{fig:mom}
\end{figure*}
\indent A striking result from Figure \ref{fig:unbnd_param} is that the rate at which fresh neutral gas is ionised is very similar in all simulations. We suggest here a possible explanation for this observation. For most of the duration of the simulations, the configuration of the gas can be thought of as approximating a leaky bubble whose inner surface is being continuously ionised by OB stars located near the centre of the bubble, while ionised gas continually escapes through the holes in the walls. If the bubble subtends a fraction $(1-f_{\rm leak})$ of the sky as seen from the ionising sources, the rate at which ionised gas escapes from the bubble is approximately
\begin{eqnarray}
\dot{M}_{\rm leak}\approx4\pi R_{\rm b}^{2}\rho_{\rm i}f_{\rm leak}c_{\rm II}
\end{eqnarray}
The density of the ionised gas can be inferred by assuming that is in pressure balance with the momentum flux of the stellar winds also emanating from the massive stars, so that
\begin{eqnarray}
\rho_{\rm i}k_{\rm B}T_{\rm i}/\mu_{\rm i}=\frac{(1-f_{\rm leak})\dot{M}_{\rm w}v_{\infty}}{(1-f_{\rm leak})4\pi R_{\rm b}^{2}}
\end{eqnarray}
If we now assert that the HII region is in a steady state, so that the rate at which material is being ionised is equal to the rate at which it is leaking from the bubble, we can combine the above two expressions to obtain
\begin{eqnarray}
\dot{M}_{\rm i}\approx f_{\rm leak}\frac{\dot{M}_{\rm w}v_{\infty}c_{\rm II}\mu_{\rm i}}{k_{\rm B}T_{\rm i}}
\end{eqnarray}
If we insert typical numbers of $f_{\rm leak}\approx0.5$, $\dot{M}_{\rm w}\sim10^{-6}$M$_{\odot}$\,yr$^{-1}$, $v_{\infty}\approx2\times10^{3}$km\,s$^{-1}$, $\mu_{\rm i}=2.3m_{\rm H}$, $T_{\rm i}=10^{4}$K, we obtain an ionisation rate of $\sim3\times10^{-4}$\,$M_{\odot}$\,yr$^{-1}$, corresponding to a mass of $\sim10^{3}$\,M$_{\odot}$ over 3\,Myr, which is close to the values we obtain from the simulations.\\
\indent \cite{2012MNRAS.427..625W} model ionising feedback in $10^{4}$\,M$_{\odot}$ non--centrally--condensed fractal clouds of initial radius 6.4\,pc and an initial mean density of $\approx 190$\,cm$^{-3}$. Their ionising sources have a common photon luminosity of $Q_{\rm H}=10^{49}$\,s$^{-1}$, comparable to the total ionising luminosities of our simulations. They find global ionisation rates rising to and plateauing at $\sim1\times10^{-3}$\,$M_{\odot}$\,yr$^{-1}$, with relatively little difference between clouds with fractal dimensions in the range 2.0--2.8. Their mean cloud density is lower than all but that in Run r07i, but even that simulation does not achieve the ionisation rates seen by \cite{2012MNRAS.427..625W}. Taken in conjunction with our findings, their results imply that a turbulent density and velocity field is more difficult to ionise than a fractal density field of comparable mean density, which may well be due to the resistance to HII region expansion offered by the ram pressures in the turbulent flows in the neutral gas.\\
\subsection{Momentum feedback from photoionisation}
\indent In \cite{2013MNRAS.436.3430D} and \cite{2014MNRAS.442..694D}, we showed that the holes though which the ionised gas leaks cause the thermal pressure in the HII region to drop to much lower values than would obtain in a sealed bubble (see figures 18 and 5 in those papers respectively, and pertinent discussion). This being the case, the principal transfer of momentum to the cold gas due to ionisation at late stages of the simulations is not due to thermal pressure of the HII region, but to the rocket effect of new gas being ionised. The transfer of momentum resulting from heating $M_{\rm i}=1\times10^{3}$\,M$_{\odot}$ to $c_{\rm II}\approx10$\,km\,s$^{-1}$ is $2M_{\rm i}c_{\rm II}\approx2\times10^{4}$\,M$_{\odot}$\,km\,s$^{-1}$ \citep{2002ApJ...566..302M}. We therefore examine in detail the uptake of momentum by the clouds.\\
\subsection{Distribution of momentum}
\indent We compute for each particle the modulus of the total momentum, the signed value of the the total radial momentum, and modulus of the azimuthal (i.e. non--radial) momentum in the clouds' centre of mass frames, and sum these quantities, distinguishing between cold neutral particles and hot ionised particles, as
\begin{eqnarray}
p_{\rm tot}=\sum_{\rm i}m_{\rm i}|{\bf v}_{\rm i}|,\\
p_{\rm rad}=\sum_{\rm i}m_{\rm i}\frac{{\bf v}_{\rm i}.{\bf r}_{\rm i}}{|{\bf r}_{\rm i}|},\\
p_{\rm azi}=\sum_{\rm i}m_{\rm i}\left({\bf v}_{\rm i}.{\bf v}_{\rm i}-\left[\frac{{\bf v}_{\rm i}.{\bf r}_{\rm i}}{|{\bf r}_{\rm i}|}\right]^{2}\right)^{\frac{1}{2}}
\end{eqnarray}
Note that, as defined, $p_{\rm tot}\neq p_{\rm rad}+p_{\rm azi}$.\\
\indent In Figure \ref{fig:mom} we plot the evolution of these quantities in time from three simulations covering the range of virial parameters under study here. Note that the control runs do not contain any hot gas. Unsurprisingly, as the virial ratio increases from run r07i through run r15l to run r23q, the momentum contained in the cold gas in the control simulation makes a larger and larger contribution to the total gas momentum in the corresponding feedback simulations, to the point where the momentum stored in the cold gas and hot gas in the dual--feedback run r23q are nearly the same at the end of the simulation.\\
\begin{figure*}
\centering
\subfloat[Interaction of bubbles driven by multiple ionising clusters in the same cloud.]{\includegraphics[width=0.32\textwidth]{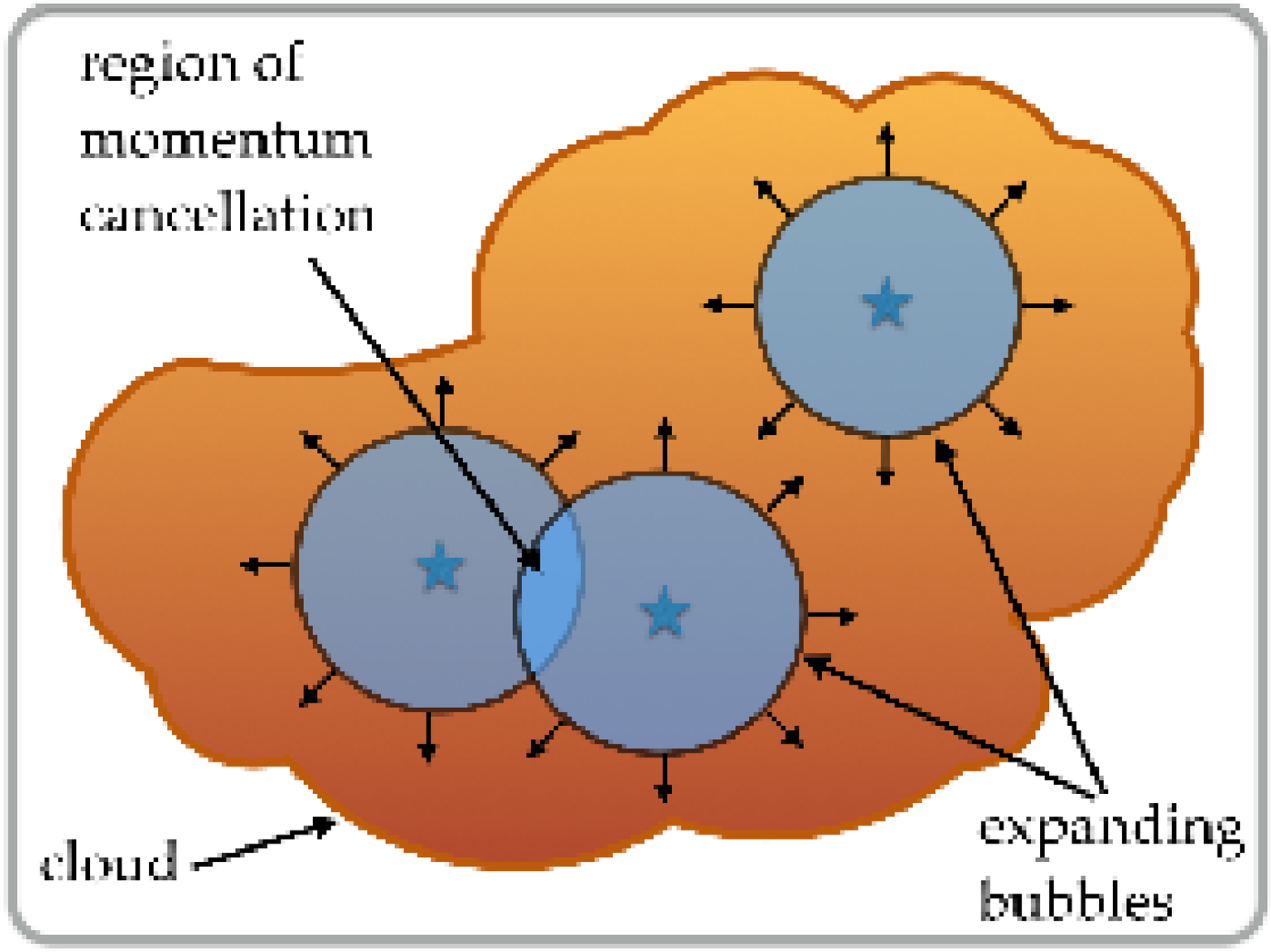}}     
     \hspace{-.01in}
\subfloat[Ionisation of a smooth shell, showing efficient momentum transfer.]{\includegraphics[width=0.32\textwidth]{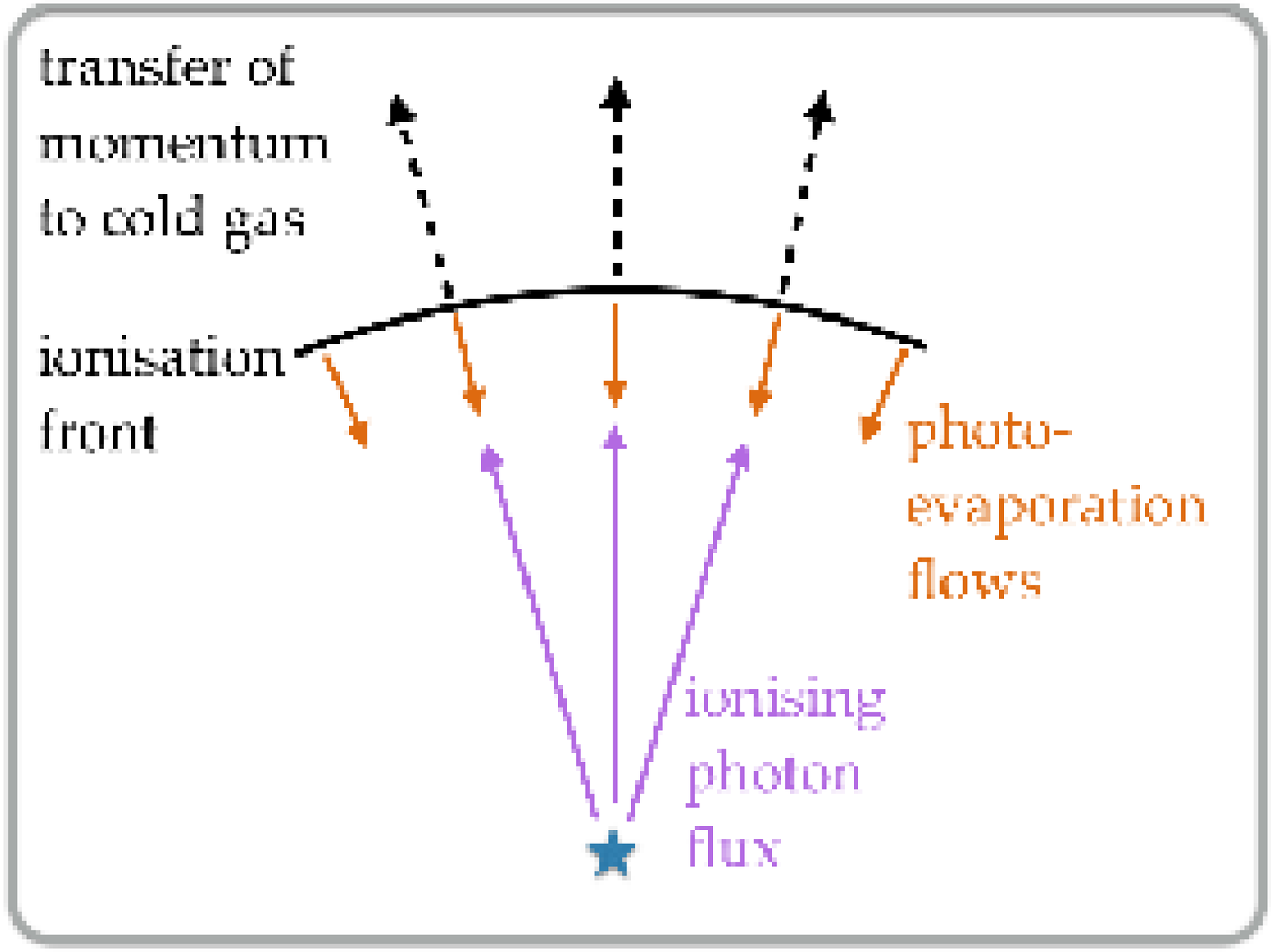}}
     \hspace{-.01in}
\subfloat[Ionisation of a shell with complex structure, illustrating partial cancellation of momentum by azimuthal flows.]{\includegraphics[width=0.32\textwidth]{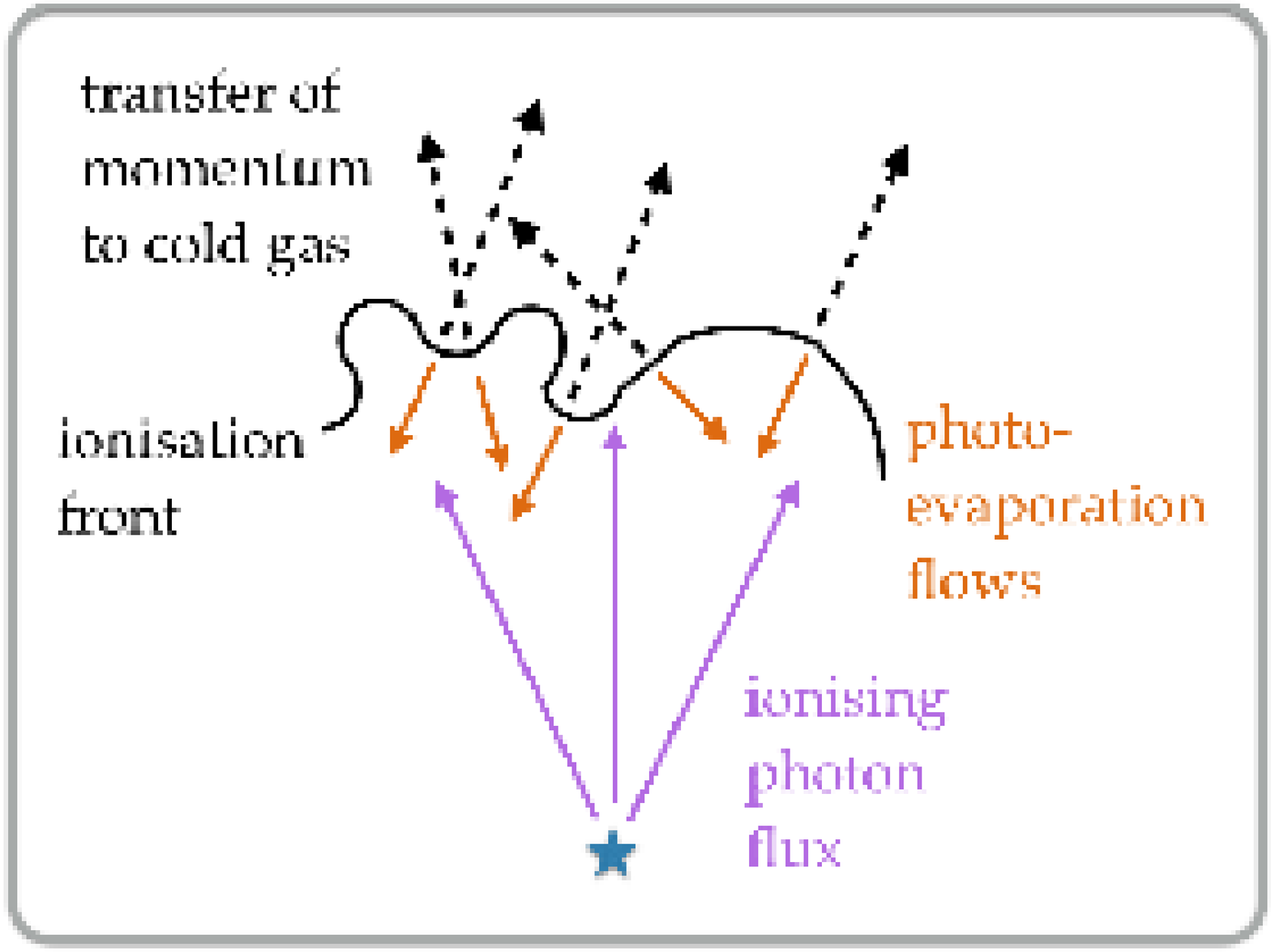}}     
\caption{Cartoon illustration of the two momentum--loss mechanisms discussed in the text.}
\label{fig:mom_cancel}
\end{figure*}
\begin{figure}
\includegraphics[width=0.48\textwidth]{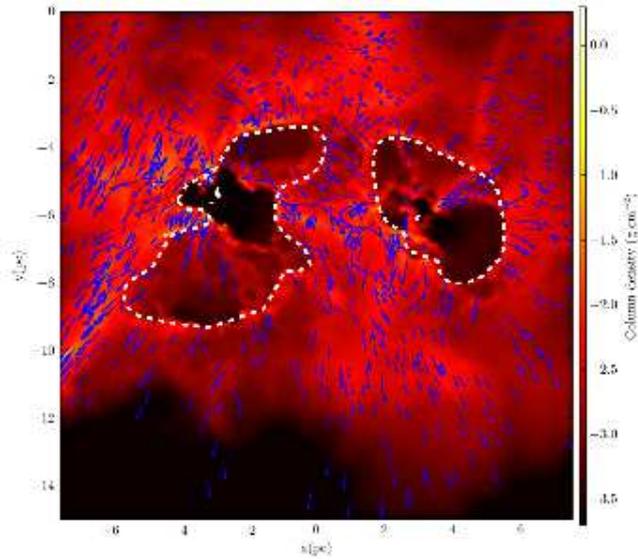}
\caption{Interaction of two bubbles at an early stage of the r23q simulation. The yellow--orange--red colour map represents gas column densities. Blue arrows show velocities of every tenth cold--gas particle. The cancellation of momentum in the region sandwiched between the two bubbles is clear from the oppositely--directed particle velocities there.}
\label{fig:coll}
\end{figure}
\begin{figure}
\includegraphics[width=0.47\textwidth]{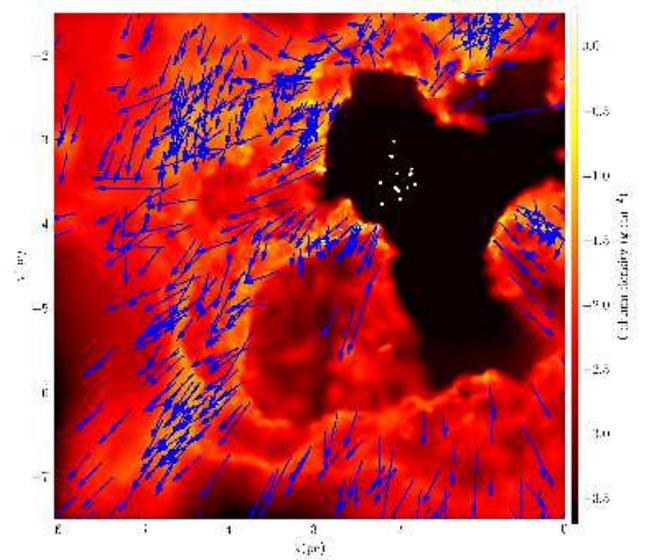}
\caption{Illustration of partial momentum cancellation due to the complex velocity field owing to photoevaporation of complex structure on the edge of an ionised bubble in the r23q simulation. The yellow--orange--red colour map represents gas column densities. Blue arrows show velocities of every tenth cold--gas particle.}
\label{fig:crossing}
\end{figure}
\subsection{Momentum loss and wastage}
\begin{figure*}
\captionsetup[subfigure]{labelformat=empty}
\centering
\subfloat[Run r07i, difference in total (left), cold gas (centre), and hot gas momentum (right) between dual--feedback and control runs.]{\includegraphics[width=0.98\textwidth]{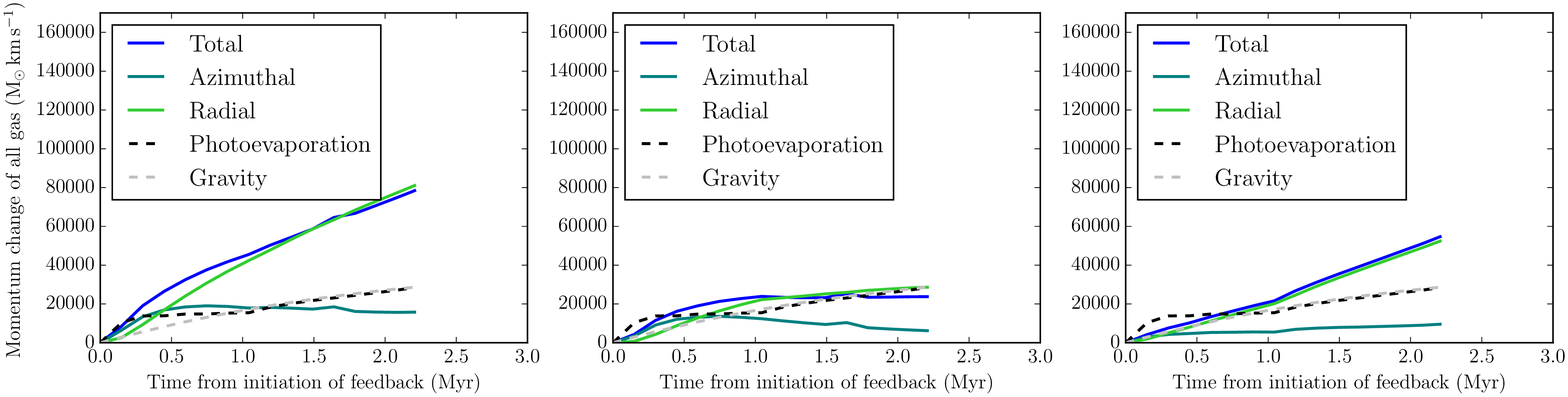}}     
     \vspace{.1in}
\subfloat[Run r15l, difference in total (left), cold gas (centre), and hot gas momentum (right) between dual--feedback and control runs.]{\includegraphics[width=0.98\textwidth]{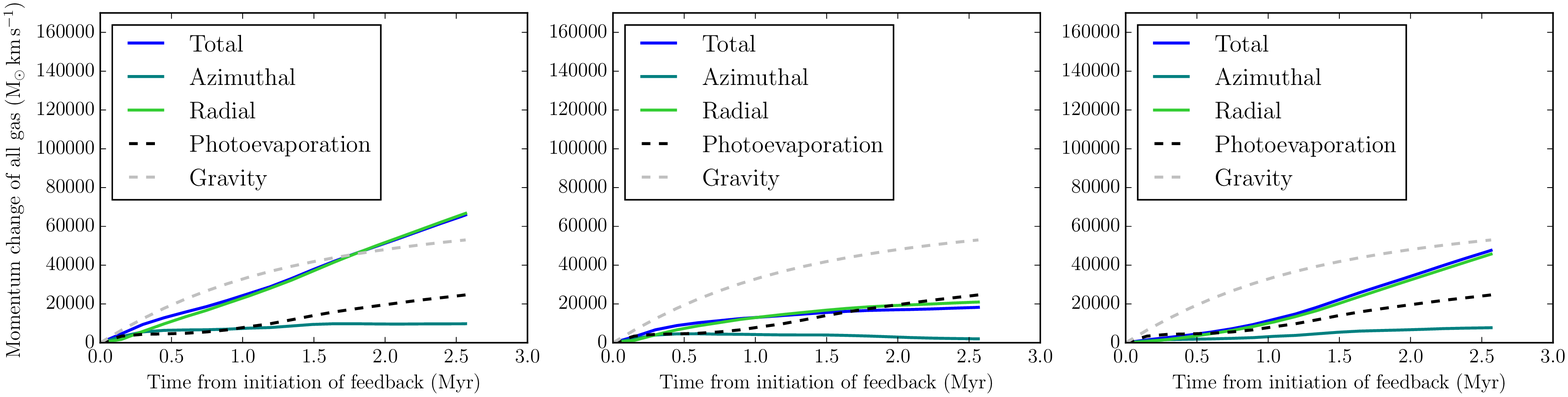}}
     \vspace{.1in}   
\subfloat[Run r23q, difference in total (left), cold gas (centre), and hot gas momentum (right) between dual--feedback and control runs.]{\includegraphics[width=0.98\textwidth]{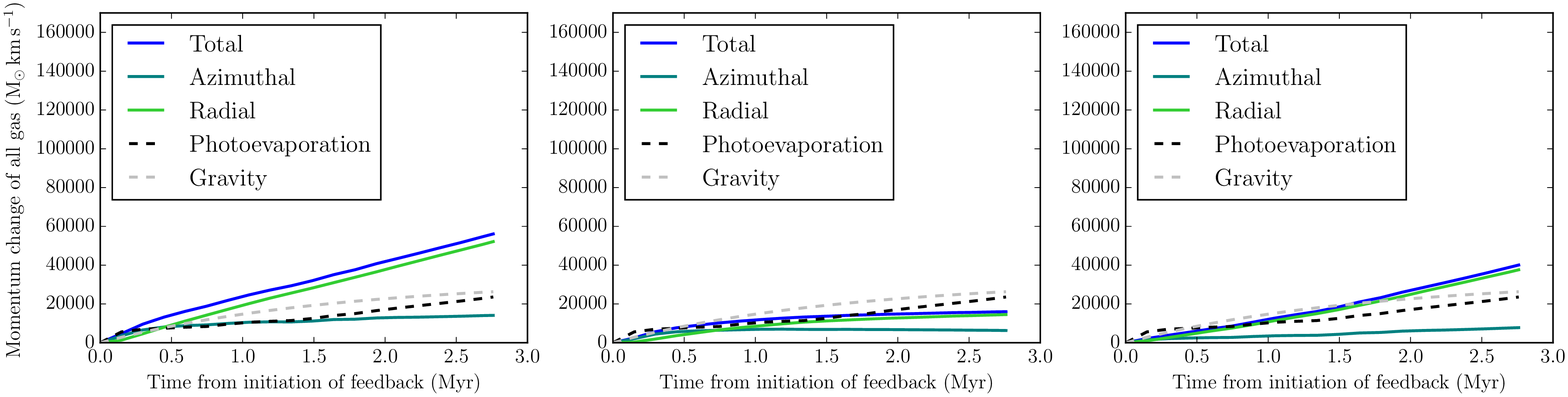}}
\caption{Time evolution of the difference in total (blue lines), radial (green lines) and azimuthal (teal lines) momenta between dual--feedback and control forks of the same three representative simulations. Black and grey dashed lines show respectively the momentum injected by photoevaporation and gravity, as described in the text.}
\label{fig:mom_change}
\end{figure*}
\indent Substantial quantities of momentum are stored in azimuthal motions. While the amount of azimuthal momentum initially also increases strongly in the dual--feedback simulations, this quantity tends to decline as the simulations progress, in contrast to the radial momentum which always increases. Given that stellar winds and HII region expansion, be it thermally or photevaporatively driven, should principally drive \textit{radial} motions, the driving of azimuthal flows demands some explanation.\\
\indent We suggest two mechanisms for driving azimuthal motions, both of which lead to momentum that could drive cloud destruction to effectively be wasted:\\ (i) the feedback sources are not all at the same location, so that some component of the flows they drive much be non--radial with respect to the cloud, and some of the momentum they inject is consumed when oppositely--directed flows collide (illustrated in the left panel of Figure \ref{fig:mom_cancel});\\
(ii) the complex structure of the gas means that momentum imparted to the cold gas by photoevaporation need not be parallel to the direction of photon propagation, which also results in the collisions of flows (illustrated by comparing the centre and right panels of Figure \ref{fig:mom_cancel}, which show the momentum transfer from a smooth and a structured shell, respectively).\\
This latter point is similar to that adduced in the driving of the ionisation--front instability \citep[e.g][]{1979ApJ...233..280G,2008ApJ...672..287W}. Here, however, the complex structures of the bubbles is more likely to be due to the complex density and velocity fields in which the HII regions are expanding.\\
\begin{figure*}
\captionsetup[subfigure]{labelformat=empty}
\centering
\subfloat[Run r07i, difference in total (left), cold gas (centre), and hot gas momentum (right) between ionisation--only and control runs.]{\includegraphics[width=0.98\textwidth]{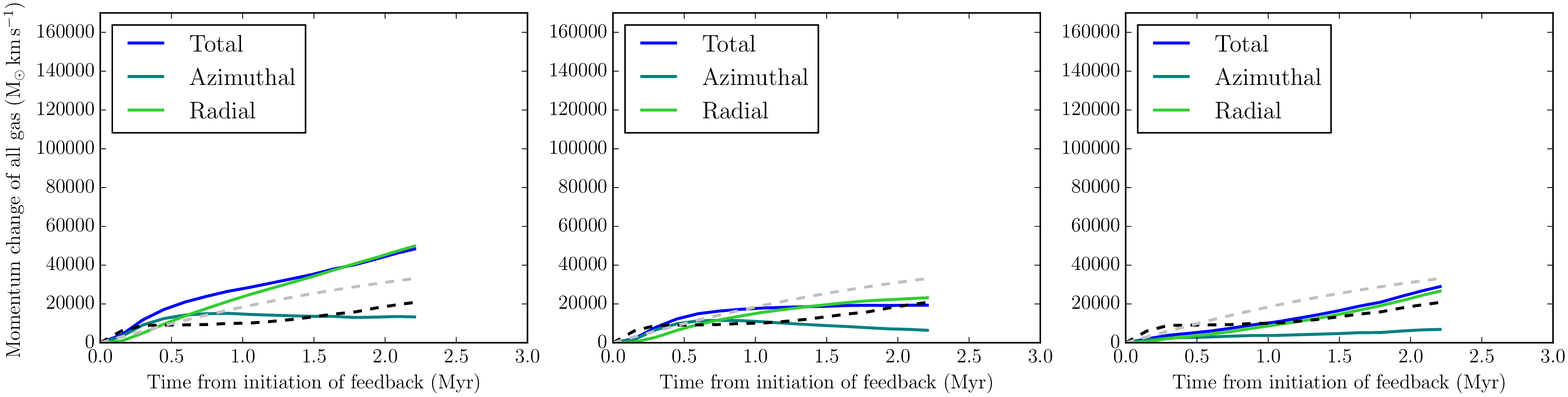}}     
\caption{Time evolution of the difference in total (blue lines), radial (cyan lines) and azimuthal (teal lines) momenta between ionisation--only and control forks of r07i simulation.}
\label{fig:mom_change_ionisation_only}
\end{figure*}
\indent In Figures \ref{fig:coll} and \ref{fig:crossing}, we show qualitatively that these two momentum--loss mechanisms do indeed operate in our simulations. Figure \ref{fig:coll} shows an early stage of the r23q simulation, where two large bubbles driven by different subclusters are interacting with each other (the evolved remains of these bubbles are visible in the bottom--centre panel of Figure \ref{fig:both_final}) The blue arrows show the velocity field of the cold gas driven by the bubbles, and it is clear that the velocities are oppositely--directly in the region squeezed between the bubbles.\\
\indent Figure \ref{fig:crossing} instead shows a region on the edge of the left--hand bubble in the same simulation at a later time, with blue arrows again depicting the velocity field of the cold gas. It is clear from the crossing of the arrows at many locations that the generally radial flows have azimuthal components which often result in their convergence, leading to partial loss of the momentum injected by photoevaporation on the inside of the bubble walls.\\
\subsection{Isolating momentum injected by feedback}
\indent As with several other quantities discussed here, it is useful to compute the differences in the values of the momenta derived above between the dual--feedback and control simulations, to isolate the changes in cloud evolution due to feedback alone. In Figure \ref{fig:mom_change}, we plot these relative quantities by subtracting the relevant momentum components of the control simulation from those of the dual--feedback simulation at each timestep, resulting in the total (blue solid lines), outward radial (green solid lines) and azimuthal (teal solid lines) momentum injected by feedback. For comparison, we also compute estimates for the cumulative momentum input from photoevaporation $p_{\rm phot}(t)$ (black dashed lines) as\\
\begin{eqnarray}
p_{\rm phot}(t)=2M_{\rm i}(t)c_{\rm II},
\end{eqnarray}
and from radial gravitational forces $p_{\rm grav}(t)$ (grey dashed lines) as
\begin{eqnarray}
p_{\rm grav}(t)=\int_{0}^{t}\frac{GM_{\rm hm}(t')^{2}}{R_{\rm hm}(t')^{2}}{\rm d}t',
\end{eqnarray}
where $M_{\rm hm}$ is half the total mass of stars and cold gas, $R_{\rm hm}$ is the radius from the system centre of mass containing $M_{\rm hm}$, and $t=0$ is the instant when feedback begins.\\
\indent Figure \ref{fig:mom_change} confirms that a large fraction of the momentum content of the dual--feedback simulations is stored in the expanding hot gas, and shows that the momentum so stored exceeds that taken up by the cold gas by factors of a few. The momentum contained in the expanding ionised gas exceeds that injected by photoevaporation, implying that the hot gas gains momentum from additional sources. The obvious candidates are thermal pressure forces, and momentum input from the stellar winds, which impact the hot gas lining the bubble interiors. We confirm that this is likely the case in Figure \ref{fig:mom_change_ionisation_only}, which depicts the evolution of the momentum--differences in the \textit{ionisation--only} r07i calculation. The momentum of the hot gas is clearly much more strongly influenced by the winds than that of the cold gas, so that the main effect of the winds is to help flush hot gas out of the cloud interiors.\\
\begin{figure*}
\captionsetup[subfigure]{labelformat=empty}
\centering
\subfloat[r07i]{\includegraphics[width=0.32\textwidth]{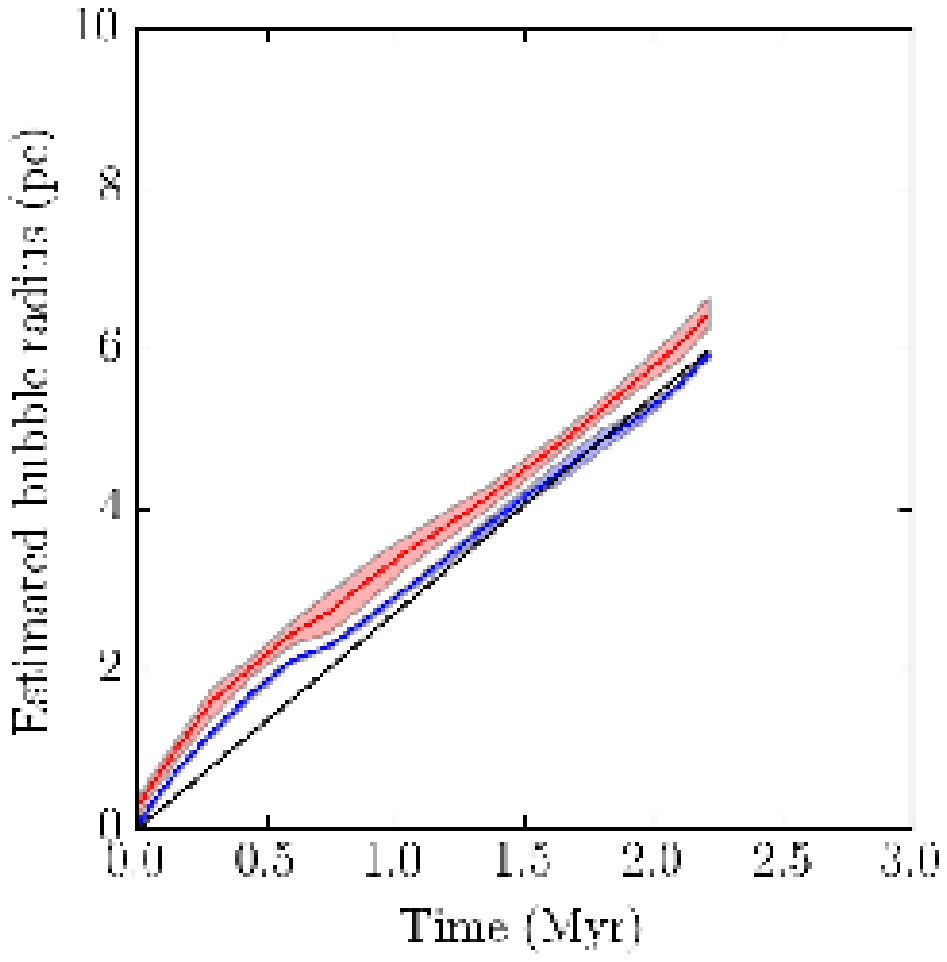}}     
     \hspace{.1in}
\subfloat[r07j]{\includegraphics[width=0.32\textwidth]{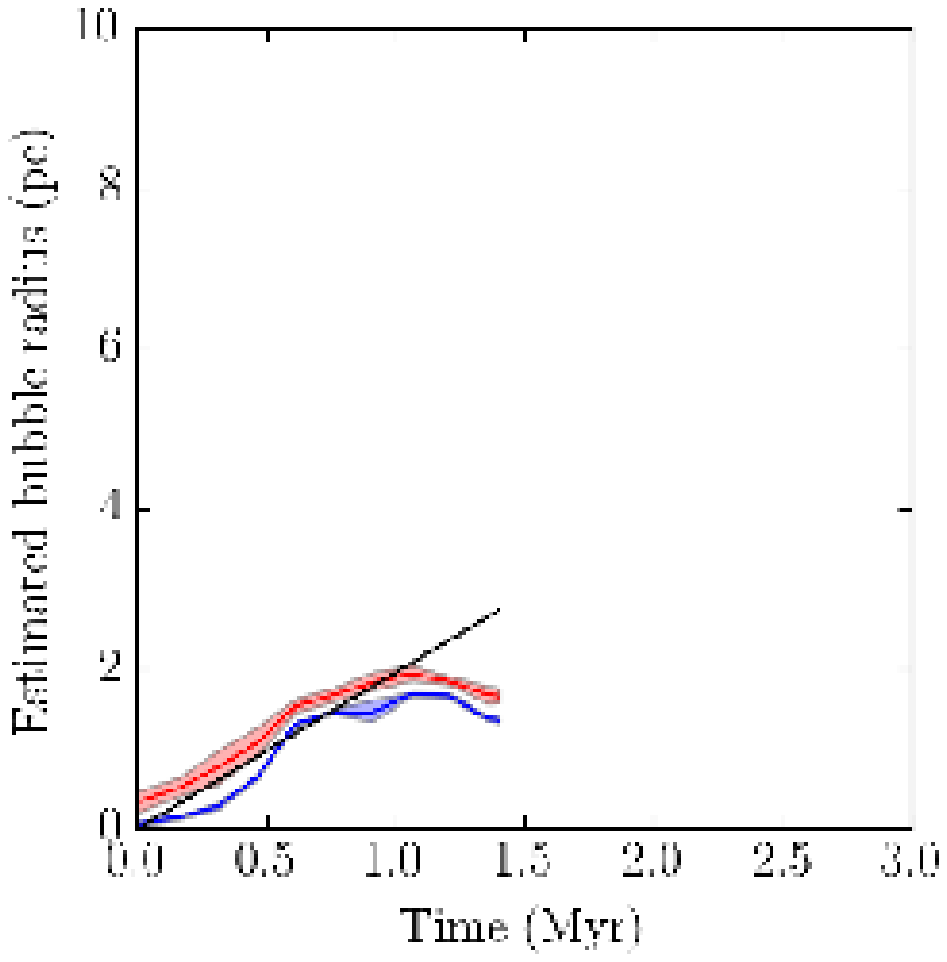}}
     \hspace{.1in}   
\subfloat[r11o]{\includegraphics[width=0.32\textwidth]{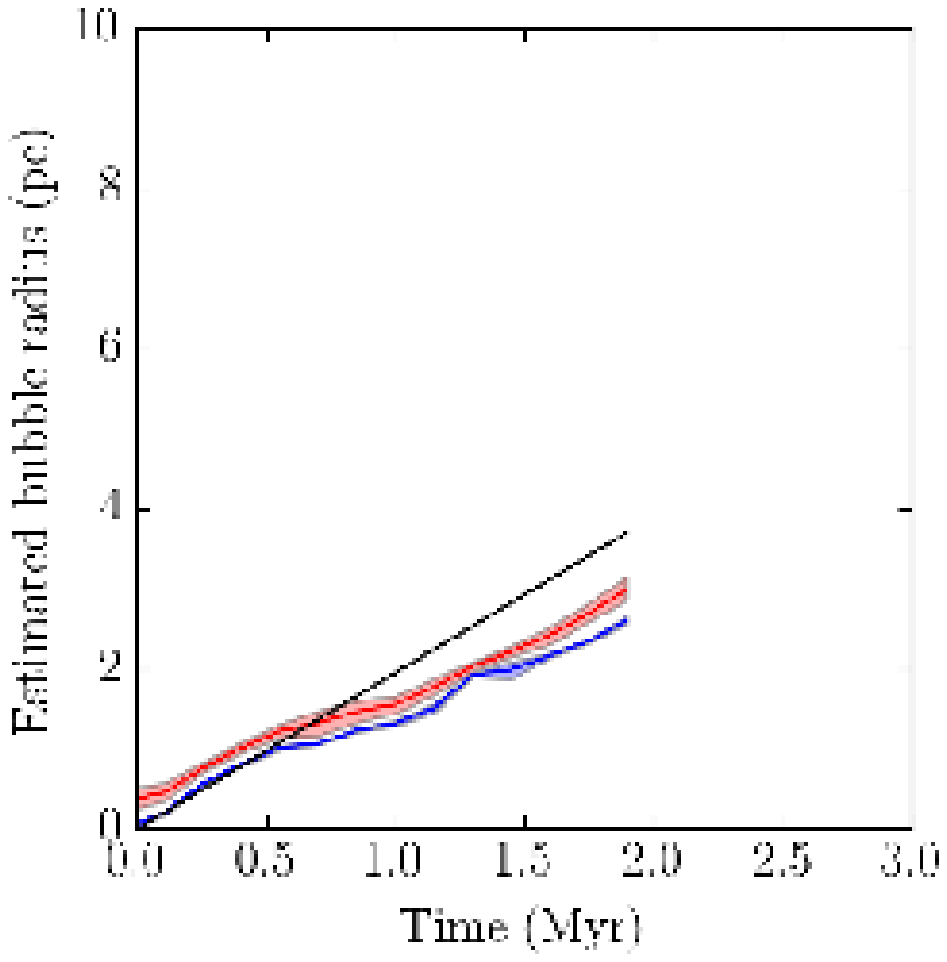}}
\vspace{0.1in}
\subfloat[r15m]{\includegraphics[width=0.32\textwidth]{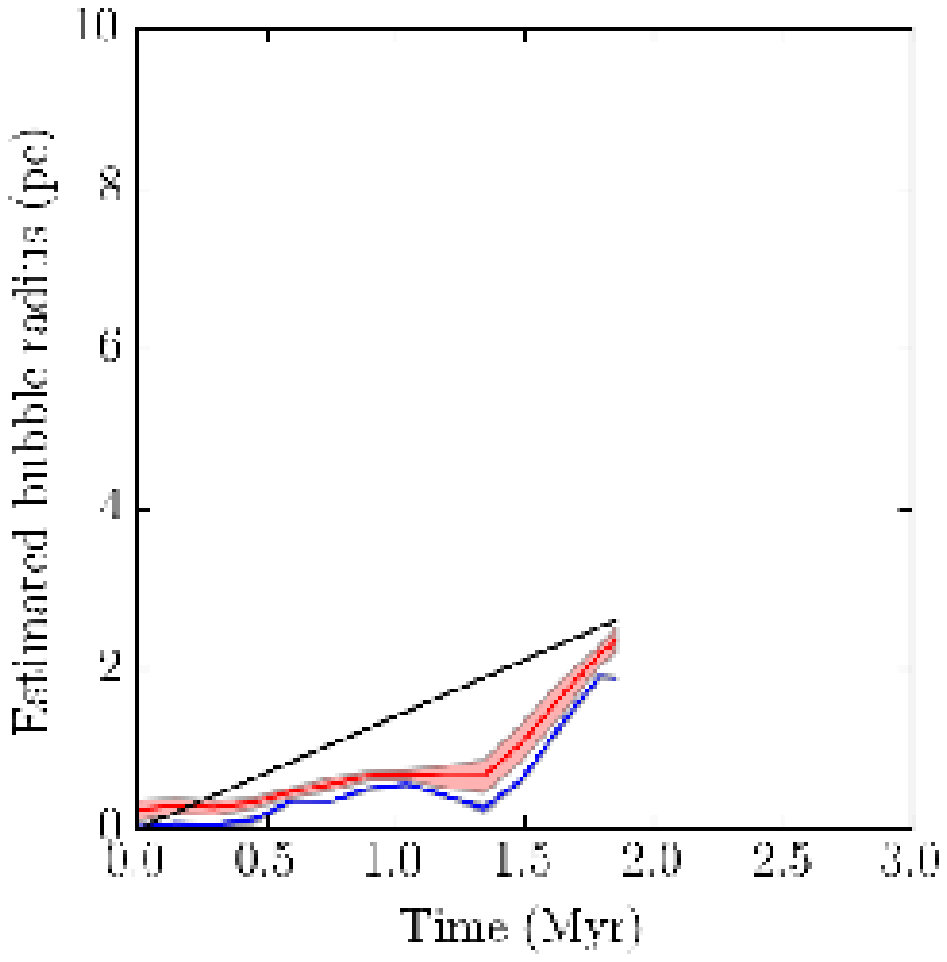}}     
     \hspace{.1in}
\subfloat[r15l]{\includegraphics[width=0.32\textwidth]{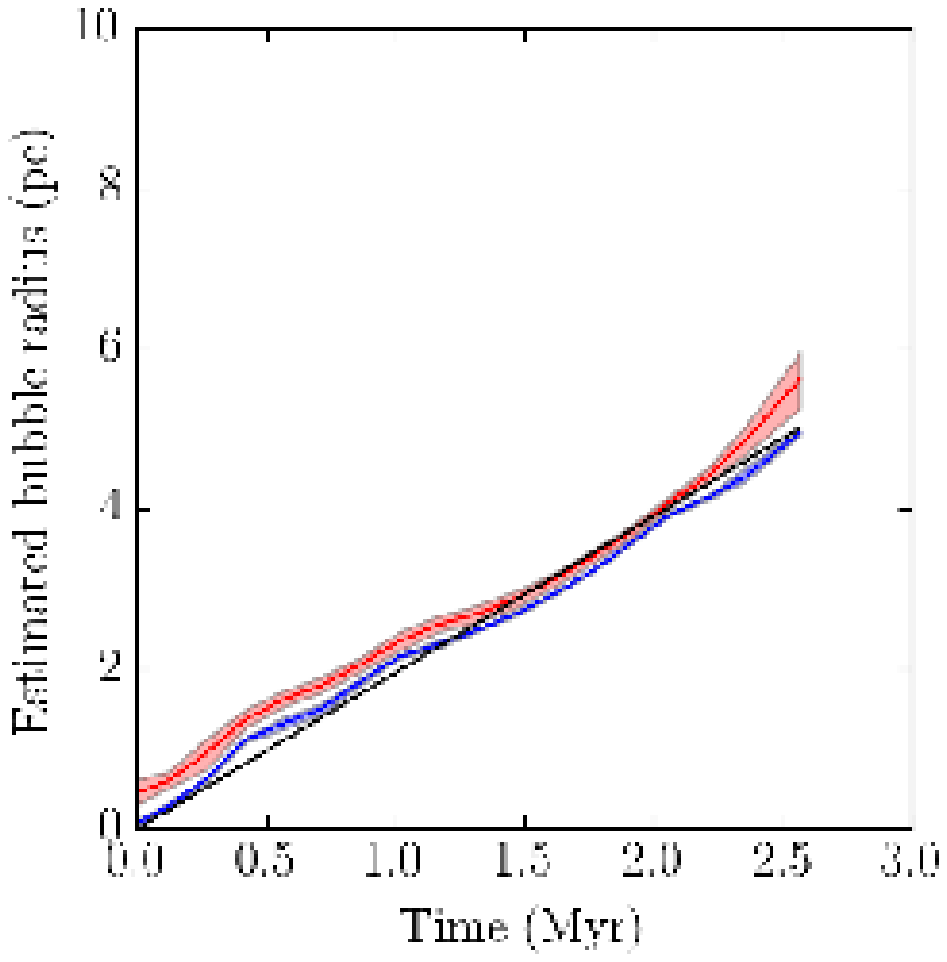}}
     \hspace{.1in}   
\subfloat[r19t]{\includegraphics[width=0.32\textwidth]{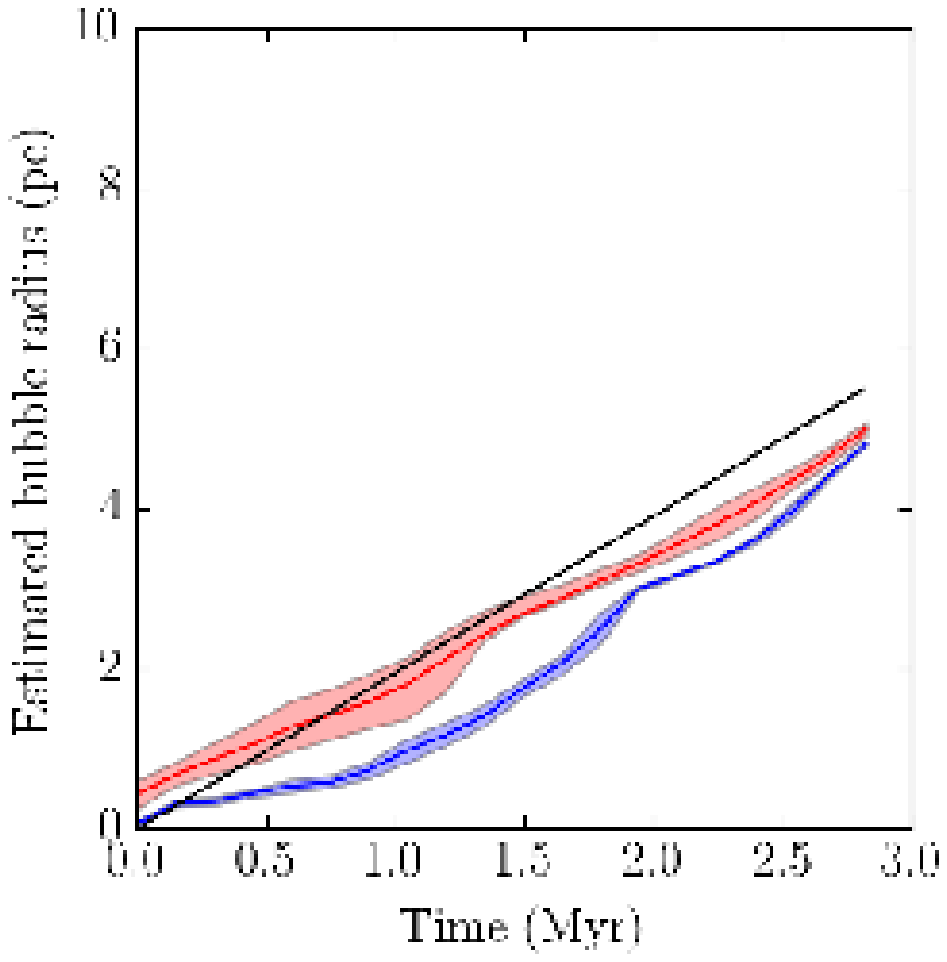}}
\vspace{0.1in}
\subfloat[r19s]{\includegraphics[width=0.32\textwidth]{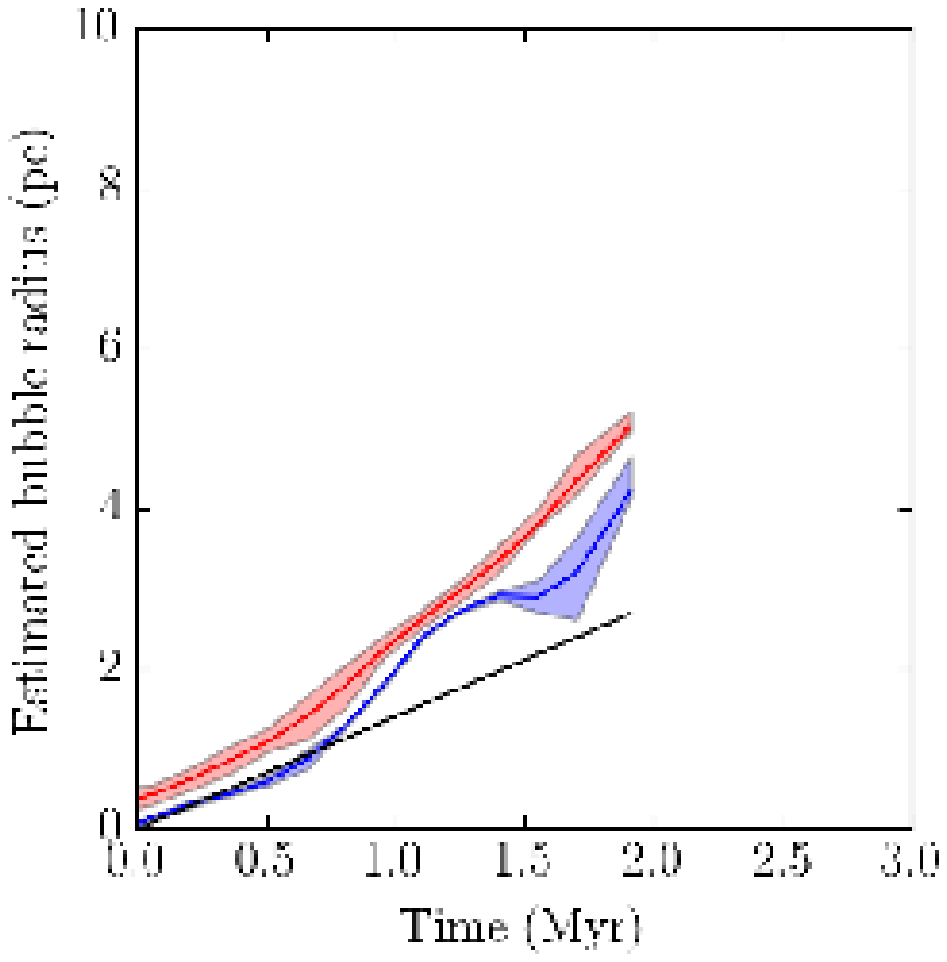}}     
     \hspace{.1in}
\subfloat[r23q]{\includegraphics[width=0.32\textwidth]{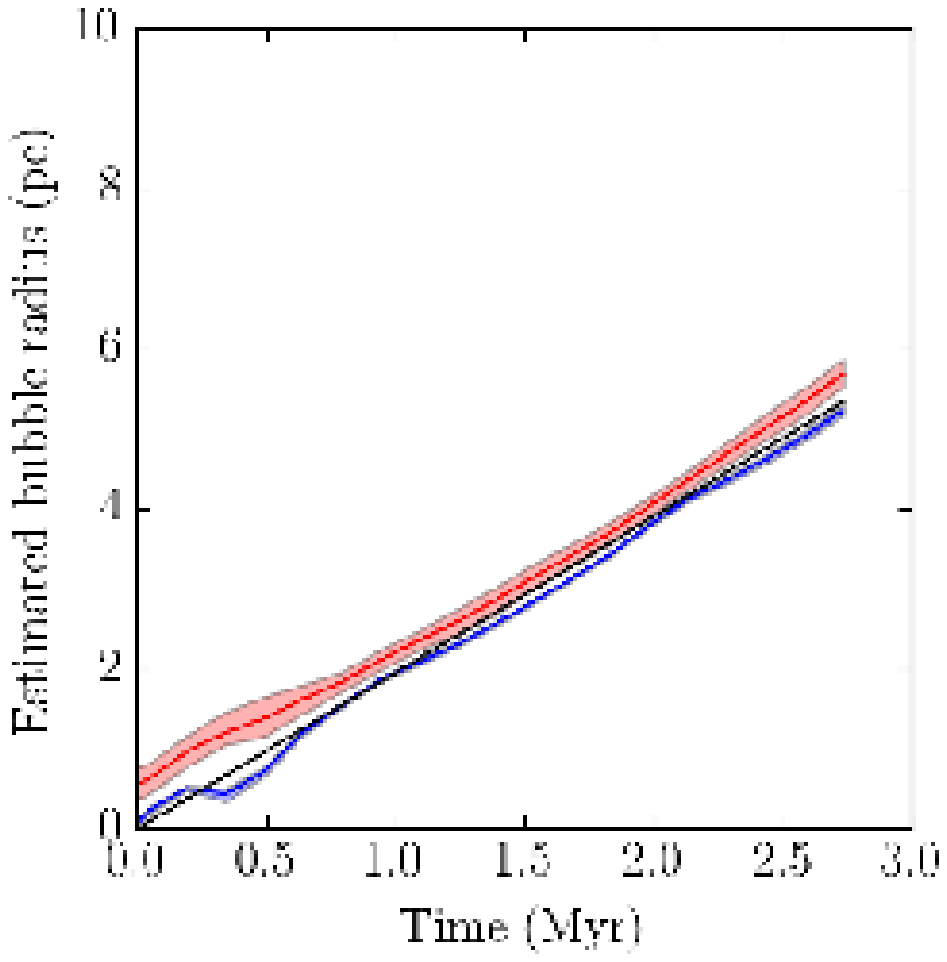}}
     \hspace{.1in}   
\subfloat[r23p]{\includegraphics[width=0.32\textwidth]{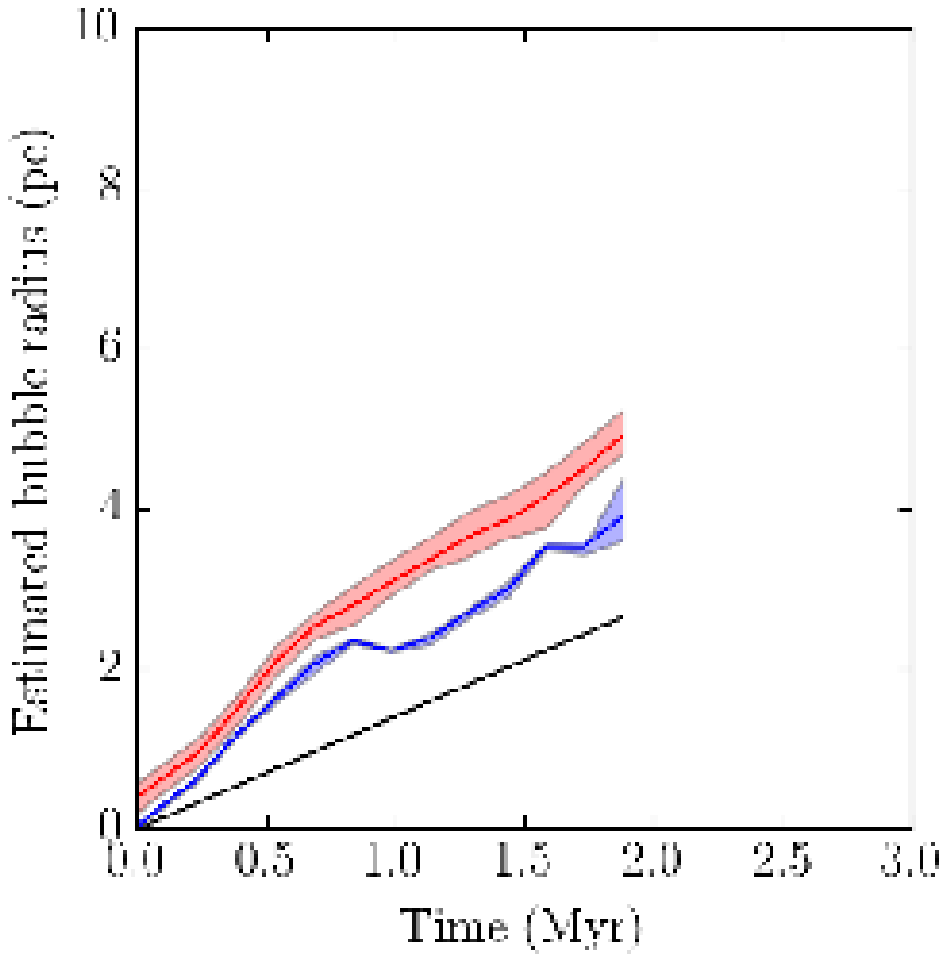}}
\caption{Comparison of the bubble expansion derived from the simple model of momentum input from photoevaporation $R(t)=k_{\rm PE}t^{n_{\rm PE}}$ (black lines) with the expansion of the $N$ nearest cold gas particles to the most massive stars (red: $N=10 000$; blue: $N=1 000$, coloured regions showing interquartile ranges) for each simulated cloud.}
\label{fig:wind_fit}
\end{figure*}
\indent Much of the momentum which has found its way into the simulations is thus contained in hot gas which has expanded to large radii and is no longer able to influence the evolution of the dense cold gas. The momentum of the cold gas at later times is instead better traced by the momentum input from photoevaporation, implying that this is indeed the main mechanism by which feedback influences the cold gas. However, at earlier times, the cold gas momentum does exceed $p_{\rm phot}(t)$, indicating that thermal pressure does contribute to the cold gas momentum budget at early times when the bubbles are still partially sealed. We find that the momentum uptake in the cold gas in these simulations (and the others, which are not shown in the interests of saving space) does indeed reach a common value of $\approx2\times10^{4}$\,M$_{\odot}$\,km\,s$^{-1}$, as suggested above by multiplying the typical mass of ionised gas by twice the ionised sound speed.\\
\indent We also see that $p_{\rm grav}$, while close to $p_{\rm phot}(t)$ in Runs r07i and r23q, is substantially larger in Run r15l (and in other simulations, particularly r07j and r15m, where the discrepancy is even greater). At first sight, this implies that momentum injected by gravitational forces in some of the simulations should completely overwhelm the effects of feedback. However, while the clouds for which this is case have lower unbound mass fractions, these fractions are still substantial ($>20$\% in all cases), so that the fact that $p_{\rm grav}>p_{\rm phot}(t)$ does not imply that feedback is unable to unbind any material from a cloud. The reason that this criterion is unreliable is that both gravity and photoevaporation inject momentum in a highly non--uniform and in homogenous manner and comparing the gross quantities of momentum input can be misleading.\\
\indent It is instructive to compare the total estimated momentum input from photoionisation with that from other feedback mechanisms. Winds are included in these simulations in a simple way as an injection of momentum. The most massive stars in our simulations have wind mass loss rates $\dot{M}_{\rm w}\approx1\times10^{-6}$\,M$_{\odot}$\,yr$^{-1}$, and terminal velocity $v_{\infty}\approx2\times10^{3}$\,km\,s$^{-1}$. Over 3\,Myr, the momentum injected by such a star is $\approx6\times10^{3}$\,M$_{\odot}$\,km\,s$^{-1}$, a factor of a few less than that introduced by photoevaporation.\\
\indent We have not included radiation pressure in our simulations, but these same massive stars have bolometric luminosities of $L_{*}\sim10^{5}$\,L$_{\odot}$, which would give rise to a corresponding radiative momentum flux $L_{*}/c\approx6\times10^{3}$\,M$_{\odot}$\,km\,s$^{-1}$, which is comparable to that injected by the winds, as might be expected. The momentum actually absorbed by the cloud depends critically upon the radiative trapping factor, which is still a matter of some controversy. \cite{2009ApJ...703.1352K} estimate it to be 2--3, while \cite{2015ApJ...809..187S} measure from their radiation transport calculations a value of barely over unity at most, and substantially less than that throughout most of their simulated clouds. The upper end of the range suggested by \cite{2009ApJ...703.1352K} would make radiation pressure of roughy equal importance as photoevaration, in terms of momentum deposition.\\
\indent For the sake of completeness, we also note that a supernova explosion in which 10\,M$_{\odot}$ of ejecta carry 10$^{51}$\,erg also carries $\approx3\times10^{4}$\,M$_{\odot}$\,km\,s$^{-1}$ of momentum. We conclude from the foregoing that momentum injected by photoevaporation is likely to be dominant at least until the detonation of supernovae and that, even after a few of these events, it will still constitute a major contribution to the momentum budget of GMCs which are in the process of disruption.\\ 
\subsection{A simple model of photovaporation--driven evolution}
\indent The HII regions in these clouds are under--pressured due to leakage of hot gas from the bubbles, and the principal transfer of momentum to the clouds is via photoevaporation. Over the course of the simulations, the quantities of mass ionised are remarkably similar (as shown in Figure\ref{fig:unbnd_param}, left panel) and the total momentum injected is also. One could in principle then regard the dynamical effect of photoionisation on the cold gas as similar to that of a steady momentum--driven wind.\\
\indent If one treats the clouds as swept--up shells of mass $M_{\rm s}$ expanding at $v_{\rm s}$ and driven by an injection of momentum resulting from a photevaporation flow at their inner surfaces, one may follow a similar procedure to that used in the analysis of momentum--driven winds in \cite{2013MNRAS.436.3430D} and write
\begin{eqnarray}
\frac{{\rm d}}{{\rm d}t}(M_{\rm s}v_{\rm s})=2\langle\dot{M}_{\rm i}\rangle c_{\rm II},
\label{eqn:shell_mom}
\end{eqnarray}
where $\langle\dot{M}_{\rm i}\rangle$ denotes the mean rate at which gas is ionised, and $c_{\rm II}$ is the sound speed in this material.\\
\indent If the clouds have density profiles described by $\rho(r)=\rho_{0}(r/r_{0})^{\alpha}$, the shell mass is given by 
\begin{eqnarray}
M_{\rm s}=\int_{0}^{R}4\pi r^{2}\rho(r){\rm d}r=\frac{4\pi\rho_{0}}{(3+\alpha)r_{0}^{\alpha}}R^{3+\alpha}
\end{eqnarray}
for $\alpha\geq-2$. Inserting this expression into Equation \ref{eqn:shell_mom}, setting $v_{\rm s}=\dot{R}$ and making the customary assumption \citep[e.g.][]{1999isw..book.....L} that $R(t)=k_{\rm PE}t^{n_{\rm PE}}$, $k$ being a constant and the subscript PE denoting photoevaporation, we obtain
\begin{eqnarray}
n_{\rm PE}=2/(4+\alpha),
\end{eqnarray}
and 
\begin{eqnarray}
k_{\rm PE}=\left[\frac{(4+\alpha)(3+\alpha)r_{0}^{\alpha}\langle\dot{M}_{\rm i}\rangle c_{\rm II}}{4\pi \rho_{0}}\right]^{\frac{1}{4+\alpha}}.
\end{eqnarray}
While leakage of the ionised gas from the cloud is essential to deriving this expression since it allows the assumption of momentum conservation to be made, the hydrodynamic leakage factor as defined in \cite{2014MNRAS.442..694D} does not appear in any of the expressions. Any dependence on the leakage factor would enter into the determination of the quantity $\langle\dot{M}_{\rm i}\rangle$, which is here treated as a constant.\\
\indent In order to test the fidelity of the relations derived above, we evaluate $R(t)=k_{\rm PE}t^{n_{\rm PE}}$ making the following assumptions:\\
(i) $\alpha=-2$;\\
(ii) $r_{0}=R_{0}/100$;\\
(iii) $\rho_{0}$ is set by integrating $\rho(r)$ from $r=0$ to $r=R_{0}$ and setting the total cloud mass to $M_{\rm c}=10^{4}$M$_{\odot}$;\\
(iv) $\langle\dot{M}_{\rm i}\rangle=\bar{f_{\rm i}}M_{\rm c}/t_{\rm SN}$ with $\bar{f_{\rm i}}$ the typical final global ionisation fraction, taken to be 0.1, and $t_{\rm SN}$ the timescale over which feedback acts, taken to be 3\,Myr, giving $\langle\dot{M}_{\rm i}\rangle\approx3.3\times10^{-4}$M$_{\odot}$\,yr$^{-1}$.\\
Under these assumptions, note that $R(t)\propto t$.\\
\indent To compare with the behaviour of the simulations, we identify the $N$ cold gas particles nearest the most massive star and compute their mean radius from the star and the interquartile range of this quantity. Following the time evolution of these quantities traces the expansion of the main bubble in each simulation. We do not use the ionisation fronts or the ionised gas to trace the bubble expansion because of the very irregular shape of the ionisation fronts and the leakage of ionised gas from the clouds to large radii. The choice of $N$ is not obvious, so we repeat the analysis using 0.1\% and 1\% of the $10^{6}$ particles in each simulation.\\
\indent We plot the results for all simulations in Figure \ref{fig:wind_fit}. Red coloured areas show the mean radius and its interquartile range for $N=10 000$, and blue areas the same quantities for $N=1 000$. The black solid line denotes the model. We note that the choice of $N$ does not affect the inferred bubble radius greatly. We also repeated the analysis using the median radius of the $N$ particles rather than the mean, but the resulting differences were negligible.\\
\indent Overall, the ability of the model to explain the evolution of the simulations is quite good. For most of the evolution of most of the simulations, the assertion that $R(t)\propto t$ is well reproduced, and the gradient (i.e. the value of $k_{\rm PE}$) is also well modelled in most simulations. The denser clouds -- r07j, r15m and r23p -- are those in which the agreement is poorest. There are many reasons which could be advocated for this, perhaps the most obvious being that the model is spherically symmetric and assumes that the feedback--driving stars are located at $r=0$, which is not strictly true in any simulation and is least accurate in the denser clouds which are very irregularly cleared by feedback. However, we consider that the agreement shown in Figure \ref{fig:wind_fit} establishes that the assertion that the clouds are largely impacted by momentum imparted by photoevaporation flows is a reasonable one. We do not propose to try to make the model more complex in the hope of achieving closer congruence with the SPH simulations.\\
\subsection{Observations of photoevaporation}
\indent These simulations suggest that photoevaporation is a process of prime importance in the disruption of molecular clouds and it is of paramount importance to test this assertion observationally by measuring photoevaporation rates directly. This can be done by measuring the electron density close to the ionisation front, in order to trace the most recently ionised gas, assuming that the gas streams away at the typical ionised sound speed of $\approx10$\,km\,s$^{-1}$, and multiplying by a suitably--measured area. The advent of advanced integral--field units has recently made it possible to measure photoevaporation rates over significant areas of star--forming regions. \cite{2015MNRAS.450.1057M} and McLeod et al. 2016 (submitted) recently employed MUSE at the VLT to measure the electron density via the $S^{23}$ sulphur line ratio for pillar structures respectively in M16, the Carina Nebula and NGC 3603. Their results clearly demonstrate the importance of photoevaporation in sculpting the remains of the natal molecular clouds in these systems, and they highlight that photoevaporation rates even for the isolated portions of the clouds represented by the pillars range over four orders of magnitude.\\
\subsection{Physics missing from this work}
\indent There are several potentially important physical processes absent from the simulations presented here. Feedback from low-- and intermediate--mass stars is neglected. As pointed out by \cite{2009ApJ...703..131O}, this is the only form of feedback operative between the onset of star formation and the birth of the first OB--type stars. Roughly speaking, such feedback falls into two classes -- thermal feedback from protstellar contraction, deuterium burning and the thermalisation of gravitational potential energy, and momentum--feedback from jets. The effects of the former are largely to suppress fragmentation and control the initial mass function \citep[e.g][]{2009MNRAS.392.1363B,2009ApJ...703..131O,2010ApJ...710.1343U}. The latter form is more effective in limiting the overall growth in stellar mass \citep[e.g][]{2012ApJ...747...22H,2015MNRAS.450.4035F}. This is of particular interest in attacking the persistent problem of simulations forming stars too quickly. It is possible that including low--mass stellar feedback in addition to the forms of feedback modelled here would produce star formation rates more compatible with those observed.\\
\indent Magnetic fields are likewise not treated here. These also have several possible modes of influence. At early stages in GMC evolution, they effectively provide an additional (non--isotropic) pressure, which slows cloud collapse somewhat \citep[e.g][]{2008MNRAS.385.1820P}. Magnetic fields may then provide an additional brake on the global star formation rate which, in combination with that due to low--mass stellar feedback, may buy more time for the massive stars to actually disrupt clouds and terminate star formation.\\
\indent However, magnetic fields will also likely influence the severity of high--mass stellar feedback. \cite{2007ApJ...671..518K} and \cite{2011ApJ...729...72P} showed that smooth magnetic fields suppresses the expansion of HII regions, although \cite{2011MNRAS.414.1747A} found that a disordered field had a much more modest influence, at least during the earlier stages of HII region expansion. Perhaps of more relevance here, \cite{2012ApJ...745..158G} examined the effects of uniform fields on blister--type HII regions or champagne flows. They found that the magnetic field reduced the hydrodynamic kinetic energy of the flow, but that the field was capable of storing a quantity of energy which far outweighed this reduction. However, it is not clear how this energy could be later tapped. The effects of magnetic fields on very leaky bubbles growing in turbulent clouds, such as those presented here, is rather unclear.\\
\indent There has been considerable recent interest in modelling radiation pressure, in particular in dense GMCs. \cite{2015ApJ...809..187S} find that radiation pressure from star clusters formed in clouds of 10$^{6}$\,M$_{\odot}$ and 10\,pc in radius, corresponding to a mean density of $\approx 5\times10^{3}$\,cm$^{-3}$. They find that radiation pressure is able to terminate star formation and disrupt the clouds. Their models are less dense than our Runs r15l, r19t and r23p, implying that these simulations might be sensitive to radiation pressure, although the simple analytic calculation given above implied that the radiative trapping factor would need to be $>3$ for radiation pressure to be more importance than photoevaporation, whereas \cite{2015ApJ...809..187S} find that the trapping factor is rarely in excess of unity. The importance of radiation pressure on low--mass clouds such as those examined here is thus not obvious.\\
\indent The expansion of HII regions and wind bubbles, particularly in declining density profiles, is likely to be subject to several hydrodynamic instabilities, such as the Rayleigh--Taylor, Vishniac \citep{1983ApJ...274..152V}, Franco \citep{1990ApJ...349..126F}, and thin--shell \citep{1994ApJ...427..384E} instabilities. These may be poorly resolved in SPH, since small--scale perturbations are smeared out by the smoothing process, retarding the growth of the instabilities.\\
\indent \cite{2009MNRAS.398.1537D} demonstrated, using detailed comparisons between the SPH code used for the calculations presented here, the FLASH \citep{2000ApJS..131..273F} grid code, and the analysis of \citep{1994ApJ...427..384E}, that SPH codes in fact are able to model the thin--shell instability. However, in \cite{2013MNRAS.436.3430D} (section 5.3) we showed that the steep radial density profiles of our clouds is likely to suppress this instability.\\
\indent Of more relevance here, \cite{1990ApJ...349..126F} showed that HII region expansion in a density profile steeper than $\rho(r)\propto r^{-3/2}$ is guaranteed to be unstable because the expansion is unable to sweep up enough material to contain the ionisation front. The azimuthally--averaged density profiles of our model clouds are in the range $\rho(r)\propto r^{-3/2--2}$. Even if the clouds were initially smooth, one would thus expect them to be susceptible to this instability. However, the clouds are strongly inhomogeneous before the onset of feedback. This is only likely to make the Franco instability more violent, since it offers directions in which the ionisation front is able to accelerate more or less than average. The front therefore tends to burst out in directions in which the local radial density profile is steepest, or in which there is the least material to sweep up.\\
\indent Our simulations do clearly exhibit this phenomenon, as shown by the early HII region morphologies depicted in Figure 1 of \cite{2014MNRAS.442..694D}. The ionisation fronts rapidly escape to large radii and ionised gas penetrates large fractions of the cloud volumes. The rapidity of this behaviour is likely to lessen the importance of other types of instabilities by washing out their effects. The existence of ionised gas throughout much of the cloud volume allows radiative feedback to influence the dynamics of a much large fraction of the cold gas in the clouds than would be allowed by the slower expansion of confined swept--up shells. \cite{2015MNRAS.447.1341B} showed that the velocity fields in the photoionised clouds do indeed possess some characteristics of turbulence, consistent with driving on scales of several pc.\\
\section{Conclusions}
We have presented simulations of the effects of ionisation and wind feedback on a set of nine 10$^{4}$\,M$_{\odot}$ turbulent molecular clouds covering a range of initial virial ratios between 0.7 and 2.3 in order to evaluate the influence of the cloud virial ratio on the evolution of the clouds. Our main conclusions are as follows:\\
(i) Star formation efficiency rates measured absolutely and per freefall time vary considerably amongst the model clouds, being highest for clouds with the shortest freefall times and lowest virial ratios. The star formation rate per freefall time in particular declines linearly with increasing virial parameter. Feedback, in the main, does not change these relations substantially, but lowers the star formation efficiency rates by similar modest factors of 30--50\% across the parameter space. This results in seven of the nine clouds exhibiting $SFER_{\rm ff}<0.05$.\\
(ii) Feedback fails to terminate star formation in any cloud, although several clouds sport (sub)clusters which are free of gas and not accreting, and where star formation has been locally stopped.\\
(iii) The fractions of the cloud masses unbound by feedback vary  somewhat more across the parameter space, being confined to the range $20-60\%$, clouds with the lowest escape velocities being those most strongly affected, in agreement with our previous studies.\\
(iv) Depressurisation of the HII regions in these low--mass clouds by hydrodynamic leakage \citep{2013MNRAS.436.3430D,2014MNRAS.442..694D} allows the influence of photoionisation to be adequately explained as being primarily doe to momentum deposition via the rocket effect, despite the fact that HII regions are generally regarding as a form of \textit{thermal} feedback. Momentum deposition is non--uniform and its efficiency is somewhat limited by interactions of separate bubbles, and by crossing of flows at irregular ionisation fronts, leading to the deposition of momentum in azimuthal directions, increasing the probability of momentum cancellation in opposing flows.\\
(v) Since all the clouds exhibit similar time--averaged ionisation rates, the momentum injection rates are approximately constant and cloud evolution may be modelled as being due to momentum--driven bubbles expanding in power--law density profiles. The density profiles of our model clouds are well approximated by $\rho\propto r^{-2}$, resulting the expansion of the bubbles being linear in time.\\
\section{Acknowledgements}
This research was supported by the DFG cluster of excellence `Origin and Structure of the Universe' (JED).

\bibliography{myrefs}

\label{lastpage}

\end{document}